\documentclass[]{raa}

\usepackage{natbib}
\usepackage{amssymb,amsmath}
\usepackage{graphicx,times}
\bibpunct{(}{)}{;}{a}{}{,}
\usepackage[pdftex,a4paper=true,pagebackref=true]{hyperref}
\hypersetup{colorlinks = true, linkcolor = green, anchorcolor = red, citecolor = blue, filecolor = red, pagecolor = red, urlcolor = red}
\usepackage{multirow}

\newcommand{\reply}[1]{{\color{black}#1}} %reply to referee

\begin{document}
    \title{Systematic errors induced by the elliptical power-law model \\in galaxy-galaxy strong lens modeling}
    \volnopage{Vol.0 (20xx) No.0, 000--000}      %%preserved for Editor. DOn't remove!
    \setcounter{page}{1}          %%starting page, preserved for Editor. DOn't remove!
    \author{
        Xiaoyue Cao\inst{1,2},
        Ran Li\inst{1,2},
        J. W. Nightingale\inst{3},
        Richard Massey\inst{3},
        Andrew Robertson\inst{3},
        Carlos S. Frenk\inst{3},
        Aristeidis Amvrosiadis\inst{3},
        Nicola C. Amorisco\inst{3},
        Qiuhan He\inst{3},
        Amy Etherington\inst{3},
        Shaun Cole\inst{3},
        Kai Zhu\inst{1,2}
        }
    \institute{
National Astronomical Observatories, Chinese Academy of Sciences, 20A Datun Road, Chaoyang District, Beijing 100012, China; {\it ranl@bao.ac.cn}\\
\and 
School of Astronomy and Space Science, University of Chinese Academy of Sciences, Beijing 100049, China\\
\and 
Institute for Computational Cosmology, Physics Department, Durham University, South Road, Durham, DH1 3LE, UK\\
\vs\no
{\small Received  20xx month day; accepted  20xx  month day}
}

    \abstract{
The elliptical power-law (EPL) model of the mass in a galaxy is widely used in strong
gravitational lensing analyses. However, the distribution of mass in real galaxies is more
complex. We quantify the biases due to this model mismatch by simulating and then analysing mock
{\it Hubble Space Telescope} imaging of lenses with mass distributions inferred from
SDSS-MaNGA stellar dynamics data. We find accurate recovery of source galaxy morphology,
except for a slight tendency to infer sources to be more compact than their true size. The
Einstein radius of the lens is also robustly recovered with 0.1\% accuracy, as is the global density slope, with 2.5\% relative systematic error,
compared to the 3.4\% intrinsic dispersion. However, asymmetry in real lenses also leads to a
spurious fitted `external shear' with typical strength, $\gamma_{\rm ext}=0.015$. Furthermore,
time delays inferred from lens modelling without measurements of stellar dynamics are
typically underestimated by $\sim$5\%. Using such measurements from a sub-sample of 37
lenses would bias measurements of the Hubble constant $H_0$ by $\sim$9\%. %The next generation cosmography must use more complex lens mass models.
\reply{Although this work is based on a particular set of MaNGA galaxies, and the specific value of the detected biases may change for another set of strong lenses, our results strongly suggest the next generation cosmography needs to use more complex lens mass models.}
}

    \authorrunning{Xiaoyue Cao et al.}
    \titlerunning{Systematics from power law lens models}

    \maketitle

\section{Introduction}
\label{sec:intro}
Strong gravitational lensing is a phenomenon whereby the light of a background source galaxy is highly distorted by the gravitational field of a foreground lens galaxy, such that the source is observed as multiple images or in extended arc structures. Over the last few decades, strong gravitational lensing has become a powerful tool for astronomers, for example it has been used as a ``cosmic telescope'' to observe magnified high-redshift sources \citep{Newton11,Shu16,Cornachione18,Blecher19,Ritondale19,Rizzo20,Cheng20,Marques20,Yang21}, as a probe of the mass structure and substructure of foreground lenses \citep{lmass_Treu06,lmass_Koopmans06,lmass_Gavazzi07,lmass_Bolton08,lmass_Vegetti09,lmass_Auger10_lens,lmass_Bolton12,lmass_he18,lmass_Nightingale19,lmass_he20,lmass_Du20}, as a tool to constrain the nature of dark matter \citep{DM_Mao1998,DM_Vegetti12,DM_Vegetti14,DM_RanLi16,DM_RanLi17,DM_Ritondale19,DM_Gilman19,DM_Gilman20,DM_He20,Enzi20}, or as an independent method for cosmological parameters inference \citep{Suyu13,slope_chen19,Birrer20,wong20, Millon20}.

Currently, thousands of strong lensing candidates have been discovered in large sky surveys \citep{lens_search_1,lens_search_2,lens_search_3,lens_search_4,lens_search_5,lens_search_6, lens_search_7, lens_search_8, lens_search_9, lens_search_10, lens_search_11, lens_search_12}. When high-resolution imaging is also available, it is possible to pin down the lens potential at the location of the multiple images to within a few percent \citep{lmass_percent_1, lmass_percent_2}; infer the Hubble constant with $\sim 7\%$ precision \citep{Suyu10,Suyu13}; detect dark matter subhaloes of masses $\sim 10^8$--$10^9$~M$_{\odot}$  \citep{Vegetti10,Vegetti12}; and resolve $\sim$100\,pc morphological structures in high redshift ($z \gtrsim 2$) source galaxies \citep{Shu16,Ritondale19}.

A key step towards all these scientific goals is to infer an accurate model of the lens mass distribution. This is challenging for a number of reasons, for example, the `source-position transformation' (SPT), whereby different lens mass models can produce almost the same lensed observations by simultaneously adjusting the distribution of lens mass and source light \citep{Schneider_Sluse_14_SPT}. A well-known case of the SPT is the `mass-sheet transformation' \citep[MST]{Falco1985}, in which a rescaling of the lens' surface mass density by a factor $\lambda$, the simultaneous addition of a constant mass sheet with convergence ($1 - \lambda$), and a rescaling of the angular size of the source by a factor of $\lambda$, leaves the resulting lensed image unchanged except for a time delay. \citet{Schneider_Sluse_13} show an example of the SPT transformation, where an elliptical power-law (EPL) mass model is fitted to a mock lens composed of a Hernquist stellar mass profile \citep{Hernquist1990} and a generalized NFW \citep{nfw_profile,gnfw_Zhao1996,gnfw_Wyithe,Cappellari13} dark matter mass profile. Although the power-law mass model provides a good fit to the position of the lensed images and the central velocity dispersion of the lens galaxy, the inferred density profile and time delays are significantly offset from the truth due to the mismatch between the EPL and the more complex input lens mass distribution.

Many studies use the EPL mass model to describe the density profiles of lens galaxies when modeling galaxy-galaxy strong lenses. In reality, the structure of the lens galaxies is more complex and may not exactly follow the power-law radial profile or conform to assumptions like perfect elliptical symmetry. This simplified mass model will therefore introduce  systematic errors into the lens modeling. Although \citet{Suyu09} found that for the mass profile of lens system, b1608+656, the deviation from the EPL model is less than $\sim$2\%, \citet{Schneider_Sluse_13} argue that even if the true mass profile deviates only slightly from a power law, it may have an impact on the inferred time delay of over $\sim$10\%.

To investigate biases that could be introduced when using a simple parameteric model to describe the mass distribution of (presumably complicated) observed lenses, one approach is to generate mock lenses using galaxies from cosmological hydrodynamic simulations and then measuring the systematic error caused by fitting a parametric model (such as the EPL model), by comparing the results with the ground truth values \citep{Mukherjee18,Enzi20,Ding21}. A limitation of this method is that cosmological simulations may lack sufficient resolution to resolve the internal structure of the lens galaxy, with the softening length of these simulations being usually between $0.3-1$ kpc \citep{hydro_sim_eagle1,hydro_sim_eagle2,hydro_sim_Illustris1,hydro_sim_Illustris2,hydro_sim_Illustris3,hydro_sim_TNG1,hydro_sim_TNG2,hydro_sim_TNG3,hydro_sim_TNG4,hydro_sim_TNG5}. Works that simulate lenses in this way have noted density cores at the centre of the simulated lenses that are not consistent with observed strong lenses \citep{central_image_Rusin01,central_image_Keeton03,central_image_Winn04,central_image_Boyce06,central_image_Zhang07,central_image_Quinn16}.

Moreover, recent observations of stellar kinematics show that the hydrodynamic simulations may underestimate the average density slope within the half-light radius of Early Type Galaxies (ETGs). \citet{Ran19} analyzed the average inner density slope of more than 2000 galaxies in the SDSS-IV MaNGA survey, using the mass reconstructions given by the Jeans Anisotropic Multi Gaussian expansion (JAM) dynamics modeling from \citet{Hongyu18}. They showed that hydrodynamic simulations underestimate the average slope by about $10\%$ to $20\%$ for massive ETGs.

In this work, instead of using ETGs from cosmological hydrodynamic simulations to make mock lens samples, we select 50 ETGs that are similar to SLACS \reply{\citep[The Sloan Lens ACS Survey,][]{Bolton06}} lenses in terms of mass and morphology from the IFU observations of the MaNGA project and generate mock lenses based on the dynamical mass maps derived by the JAM method. We anticipate that these mock lenses contain many of the complexities found in the mass distributions of real galaxies (e.g. non-symmetric mass structures) which the much simpler EPL mass profile, often assumed for lens modeling, does not. Thus, this simulated dataset will allow us to assess systematic errors in lens modeling due to assuming an unrealistic mass model. In particular, we evaluate the morphology of the source galaxy, the mass model parameters (Einstein radius, shear, inner density slope) of the lens galaxy, and the time delay measurement that is vital for studies of cosmography.

This paper is organized in the following way. In Section~\ref{sec:lens_theory_and_gen_data}, we introduce the basic theory of gravitational lensing and the generation of the mock lenses. The lens modeling methodology we use is briefly described in Section~\ref{sec:lens_model_method}. Our main results are shown in Section~\ref{sec:results_and_discusion}. A summary and conclusion are given in Section~\ref{sec:conclution}. Calculations presented in this work assume the Planck cosmology \citep{planck15}, adopting the following parameter values: $\rm H_0 = 67.7 \; km\,s^{-1}\,Mpc^{-1}$, $\rm \Omega_m = 0.307$, $\rm \Omega_\Lambda = 0.693$. Throughout this paper, each quoted parameter estimate is the median of the corresponding one-dimensional marginalized posterior, with the quoted {\it statistical error} calculated from the 16th and 84th percentiles (that is, the bounds of a 68\% credible interval). The statistical error only accounts for the contribution of the image noise, and they only apply to the idealized case where the lens model can perfectly describe the observational data \citep{Bolton_sample_08}. Model biases induced by the imperfect lens model method (such as the oversimplified EPL mass model in this work) are then described by the {\it systematic error}, which is quantified by comparing values inferred by the lens model to input true values.

\section{Lens theory and mock data generation}
\label{sec:lens_theory_and_gen_data}
Section~\ref{sec:lens_theory} lists the lensing formulae used in this paper;  Section~\ref{sec:Lensing_Degeneracy} gives an overview  of the SPT and lens modeling degeneracies; Section~\ref{sec:JAM_method} reviews the framework of the JAM method and Section~\ref{sec:gen_data} describes how we generate mock lenses that are similar to real observations.

\subsection{Lensing theory}
\label{sec:lens_theory}
Gravitational lensing can be described by the following relation,
\begin{equation}
    \boldsymbol{\beta}=\boldsymbol{\theta}-\boldsymbol{\alpha} ,
	\label{eq:3_mapping}
\end{equation}
where $\boldsymbol{\beta}$ is the source-plane position, $\boldsymbol{\theta}$ is the image-plane position of the lensed image and $\boldsymbol{\alpha}$ is the deflection angle which maps between the two coordinates and depends on the mass distribution of the lens. In galaxy-scale strong lens modeling, the lens galaxy density profile is often described by an EPL surface mass density of the form,
\begin{equation}
    \kappa(r) = \frac{2 - t}{2} \left(\frac{b}{r}\right)^{t} ,
	\label{eq:4_power_law} 
\end{equation}
where $\kappa$ is the dimensionless surface mass density (i.e. convergence), $0 < t < 2$ is the power-law density slope\footnote{This is equivalent to $1 < \langle\gamma\rangle < 3$, where $\langle\gamma\rangle$ is the 3-dimensional density slope. Throughout this work, we always report the 3-dimensional density slope, unless otherwise stated.}, $b > 0$ the scale length, and $r > 0$ the radial distance from the centre of the lens. If we let the center of the lens be the origin of the two-dimensional Cartesian coordinates, then $r(x, y) = \sqrt{x ^ 2 + y ^ 2}$, and we can use the following transformation, $r(x, y) \rightarrow r_1(x_1,y_1)$ to arrive at
\begin{equation}
\begin{aligned}
& x_1 = x\cos{\phi_m} + y\sin{\phi_m}  \\
& y_1 = -x\sin{\phi_m} + y\cos{\phi_m}
\end{aligned} 
\label{eq:5_ellip_rot}
\end{equation}
\begin{equation}
    r_1(x_1, y_1) = \sqrt{x_1^2q + y_1^2/q} ,
	\label{eq:5_ellip_trans}
\end{equation}
which brings ellipticity into Equation~\eqref{eq:4_power_law}, where $q$ is the axis ratio of the elliptical mass distribution and $\phi_m$ is the position angle that is counterclockwise from the x-axis to the semi-major axis ($x_1$) of the mass distribution. The deflection angle, $\boldsymbol{\alpha}$, under the EPL mass model can be calculated efficiently using the hypergeometric function \citep{Tessore15}. The environment around a lens galaxy (e.g. mass structures along the line-of-sight) may also contribute to the deflection angles. To first order, their contribution can be approximated as an shear term whose potential, $\psi_{\rm shear}$, can be described in polar coordinates ($r$, $\phi$) by,
\begin{equation}
    \psi_{\rm shear}(r,\phi) = -\frac{\gamma_{\rm sh}}{2} r^2 \cos{2(\phi - \phi_{\rm sh})} ,
	\label{eq:6_shear}
\end{equation}
where $\gamma_{\rm sh}$ represents the shear strength and $\phi_{\rm sh}$ is the position angle of shear measured counterclockwise from the positive $x$-axis.

For a lensed image at position $\boldsymbol{\theta}$ on the image-plane (with corresponding source-plane position $\boldsymbol{\beta}$), its light travel time relative to the unlensed case (i.e, the {\it excess time delay}) is defined as 
\begin{equation}
    t(\boldsymbol{\theta},\boldsymbol{\beta}) = \frac{D_{\rm \Delta t}}{c} \left(     \frac{(\boldsymbol{\theta}-\boldsymbol{\beta})
    ^2}{2} - \psi(\boldsymbol{\theta}) \right) .
	\label{eq:7_excess_td}
\end{equation}
$\psi(\boldsymbol{\theta})$ is the lens potential at position $\boldsymbol{\theta}$, $c$ is the speed of light, and $D_{\rm \Delta t}$ is the so-called time-delay distance \citep{Refsdal1964,lensing_textbook,Suyu10}, which is given by,
\begin{equation}
    D_{\rm \Delta t} \equiv (1+z_{\mathrm{l}})\frac{D_{\mathrm{l}} D_{\mathrm{s}}}{D_{\mathrm{ls}}} .
	\label{eq:8_td_dist}
\end{equation}
Here, $z_{\mathrm{l}}$ is the lens redshift and $D_{\mathrm{l}}$, $D_{\mathrm{s}}$, $D_{\mathrm{ls}}$ are the angular distances from the observer to the lens, the observer to the source, and between the lens and the source respectively. The relative time delay between a lensed image pairs A and B, $\rm {\Delta t}_{\rm AB}$ is given by 
\begin{equation}
    \Delta t_{AB} = \frac{D_{\rm \Delta t}}{c}  \left[ \frac{(\vec{\theta_A} - \vec{\beta})^2}{2} - \frac{(\vec{\theta_B} - \vec{\beta})^2}{2} - \psi(\vec{\theta_A)} + \psi(\vec{\theta_B}) \right]~.
\label{eq:rel_td}
\end{equation}

Starting from $\psi$, we can derive the deflection angle, $\boldsymbol{\alpha}$, and convergence, ${\kappa}$, by differentiation,
\begin{equation}
    \boldsymbol{\alpha} = {\nabla} \psi ,
	\label{eq:9_diff_alpha}
\end{equation}
\begin{equation}
    {\kappa} = \frac{1}{2}{\nabla^2} \psi .
	\label{eq:10_diff_kappa}
\end{equation}
We can also derive $\alpha$ and $\psi$ from $\kappa$ via integration,
\begin{equation}
    \boldsymbol{\alpha}(\boldsymbol{\theta}) = \frac{1}{\pi}\int_{\rm I\!R^2} d^2 \boldsymbol{\theta'}     ~\kappa(\boldsymbol{\theta'})~\frac{\boldsymbol{\theta}-\boldsymbol{\theta'}}{\vert \boldsymbol{\theta} - \boldsymbol{\theta'}     \vert^2},
	\label{eq:11_int_alpha}
\end{equation}
\begin{equation}
    \psi(\boldsymbol{\theta}) = \frac{1}{\pi} \int_{\rm I\!R^2}d^2\boldsymbol{\theta'} ~ \kappa(\boldsymbol{\theta'}) ~\text{ln}     \vert \boldsymbol{\theta}- \boldsymbol{\theta'} \vert.
	\label{eq:12_int_psi}
\end{equation}
When we generate mock lenses, we use equations~\eqref{eq:11_int_alpha} and \eqref{eq:12_int_psi} to calculate $\boldsymbol{\alpha}$ and $\psi$. To calculate the critical line of a given lens, we use the Jacobian matrix, $A(\boldsymbol{\theta})$
\begin{equation}
    A(\boldsymbol{\theta}) = \frac{\partial \boldsymbol{\beta}}{\partial \boldsymbol{\theta}} = 
    \left(\delta_\mathrm{ij} - \frac{\partial^2 \psi}{\partial \theta_\mathrm{i} \partial \theta_\mathrm{j}} \right).
	\label{eq:13_jacobian}
\end{equation}
The critical line is defined as the curve of points in the image plane which satisfy the relation $\mathrm{det}A(\boldsymbol{\theta})=0$. Due to their complex mass distributions it is difficult to give an analytic Einstein radius for the lenses simulated in this work and we therefore follow a numerical approach. First, we use the \texttt{matplotlib.pyplot.contour} module to draw the critical curve numerically (whereby the contour satisfies $\mathrm{det}A=0$). The Green's function method is then used to calculate the enclosed area of the critical curve, $S_\mathrm{crit}$. Finally, the effective Einstein radius \reply{\citep[see section 4.1,][]{Meneghetti_2013}}, $\theta_{\mathrm{E}}$, is defined according to the following formula,
\begin{equation}
    S_{\mathrm{crit}} = \pi ~ \theta_{\mathrm{E}}^2  .
	\label{eq:14_def_thetaE}
\end{equation}
\reply{In this work, we always report the effective Einstein radius for both the input data and the lens model reconstruction. People may use a different definition of Einstein radius, i.e, the ``equivalent Einstein radius'', defined as the radius of a circle (or ellipse, depends on whether the lens is circular or elliptical symmetric) within which the average convergence is 1. For simple axis-symmetric lenses, those two definitions of Einstein radius give the same result.}

We use the mass-weighted density slope \citep{Dutton14} to measure the average density slope, $\langle \gamma^{\prime} \rangle$, over a certain radial range. The global density slope in the context of strong lensing is defined as
\begin{equation}
\langle \gamma_{\mathrm{global}}' \rangle  \equiv -\frac{\int_0^{R_{\theta_{\mathrm{E}}}} 4\pi r^2 \rho(r) \frac{d\log{\rho}}{d\log{r}}dr}{\int_0^{R_{\theta_{\mathrm{E}}}} 4\pi r^2 \rho(r)dr} ,
\label{eq:15_def_slope}
\end{equation}
where $\rho(r)$ is the density at radius $r$ and $R_{\theta_{\mathrm{E}}}$ is the physical size of the Einstein radius.
The local density slope measures the mean density slope near the Einstein radius, which is defined similarly as
\begin{equation}
\langle \gamma_{\mathrm{local}}' \rangle  \equiv -\frac{\int_{R_{\rm 1}}^{R_{\rm 2}} 4\pi r^2 \rho(r) \frac{d\log{\rho}}{d\log{r}}dr}{\int_{R_{\rm 1}}^{R_{\rm 2}} 4\pi r^2 \rho(r)dr} ,
\label{eq:15_def_slope_2}
\end{equation}
where $R_{\rm 1}$ and $R_{\rm 2}$ are the inner and outer radius of the annulus which encloses the extended lensed arc. Based on the true mass profile of the mock lenes,  we calculate the ``global'' and ``local'' density slopes with the above equations to represent the true slope value of our mock data. These slope values are compared with the one reconstructed by lens modeling under the assumption of the EPL + shear mass model in Section~\ref{sec:results_and_discusion}.

%-----------------------------Lensing Degeneracy
\subsection{Lensing Degeneracy}
\label{sec:Lensing_Degeneracy}
The Source-Position Transformation (SPT), which describes an intrinsic degeneracy in lens modeling, shows that different lens mass models can produce almost the same lensing observable by adjusting the lens mass and source light in a covariant fashion \citep{Schneider_Sluse_14_SPT}. The well-known Mass-Sheet Transformation \citep[MST, a special case of SPT]{Falco1985} requires the source size\footnote{\reply{Strictly speaking, the change of source size is a direct result of rescaling the entire source-plane coordinates from ($x_s,y_s$) to ($\lambda \times x_s, \lambda \times y_s$).}}, $r_{\mathrm{e}}$, and lens convergence field, $\kappa$, to be transformed according to
\begin{equation}
    \kappa^{\prime} = \lambda \times \kappa + (1-\lambda),
	\label{eq:1_MST}
\end{equation}
and
\begin{equation}
    r_{e}^{\prime} = \lambda \times r_{e},
	\label{eq:2_MST}
\end{equation}
which keeps the lensed images invariant except for the time delay \reply{and magnification}. The relative time delay value, $\rm \Delta t_{\rm AB}$, between a pairwise lensed images A and B changes according to,
\begin{equation}
\Delta t_{\rm AB}^{\prime} = \lambda \times \Delta t_{\rm AB} .
\label{eq:3_MST}
\end{equation}

Lens modeling constrains the deflector's mass distribution by fitting to the lensed emission, which depends upon the deflection angle field in the extended arc region. For an azimuthally symmetric lens, the deflection angles are related to the mean enclosed convergence within $\theta$ ($\bar\kappa(<\theta)$) via
\begin{equation}
\alpha(\theta)={2\over \theta}\int_0^\theta
{\rm d}\theta' \;\theta'\,\kappa(\theta') = 
\bar\kappa(<\theta)\theta\;.
\label{eq:4_alpha_kappa_mean}
\end{equation}
Suppose we have a deflector with a true mass distribution $\kappa_{\mathrm{true}}$. All of the mass-sheet transformed solutions of $\kappa_{\mathrm{true}}$, which we can define as 
\begin{equation}
\kappa_\mathrm{MST} = \lambda \times \kappa_{\mathrm{true}} + (1-\lambda), 
\end{equation}
provide equally good fits to the image-based lens modeling because of the MST. When we take the EPL mass model for lens modeling, the mismatch between $\kappa_{\mathrm{true}}$ and the EPL model can often be compensated by a `pseudo internal mass sheet'\footnote{We call it the ``pseudo'' internal mass-sheet because it does not relate to any physical mass structure. It purely originates from the mass distribution mismatch between the ideal EPL model and real lenses.} ($1-\lambda_{\mathrm{pseudo}}$). More specifically, the transformed mass distribution $\rm \kappa^{\prime} = \lambda_{\mathrm{pseudo}} \times \kappa_{\mathrm{true}} + (1-\lambda_{\mathrm{pseudo}})$,
can often closely approximate an EPL model in terms of the deflection angle field (or $\bar\kappa(<\theta)$) near the Einstein ring. The selection of the EPL functional form therefore drives the lens model to infer the deflector's mass distribution as an EPL approximating $\rm \kappa^{\prime}$.

Following the discussion in \citet{Xu16, Schneider_Sluse_13}, the specific value of $\lambda_{\mathrm{pseudo}}$ can be derived analytically. We use an annular region between 0.8 and 1.2 times the Einstein radius to represent the extended-arc region\footnote{We have also tried a different radial range: [$0.5\times\theta_{\rm E}$, $1.5\times\theta_{\rm E}$]. Although the specific $\lambda_{\mathrm{pseudo}}$ value calculated from equation~\eqref{eq:mst_lambda} vary slightly, the main results in this work remain unchanged.}. The average density slope ($s$) and curvature ($\xi$) within the annular region ([$\theta_{\rm 1}$, $\theta_{\rm 2}$]) are defined as
\begin{equation}
s\equiv\frac{\ln(\bar{\kappa}_{\rm 2}/\bar{\kappa}_{\rm 1})}{\ln({\theta}_{\rm 1}/{\theta}_{\rm 2})},
\label{eq:mst_slope}
\end{equation}
\begin{equation} 
\xi\equiv\frac{\bar{\kappa}(\sqrt{\theta_{1}\theta_{2}})}
              {\sqrt{\bar{\kappa}_{\rm 1}\bar{\kappa}_{\rm 2}}},
\label{eq:mst_curvature}
\end{equation}
where $\bar{\kappa_1}$ and $\bar{\kappa_2}$ are the average convergence of the deflector within $\theta_{\rm 1}$ and $\theta_{\rm 2}$ respectively. $\xi$ is used to describe the degree to which the density profile deviates from the power law. When $\xi=1$, the deflection angle field in the annular region is closest to a power law. The $\lambda$ factor which makes the curvature of the profile $\lambda \times \bar{\kappa} + (1-\lambda)$ equal to 1 (i.e, $\xi_{\rm \lambda}=1$) is then given by
\begin{equation}
\lambda= \frac{1+d^s-2\,\xi\, d^{s/2}}{1+d^s-2\,\xi\, d^{s/2} + (\xi^2-1)\bar{\kappa}_{\rm 2}}
\label{eq:mst_lambda}
\end{equation}
where $d \equiv\theta_{\rm 1}/\theta_{\rm 2}<1$, so that $\bar{\kappa}_{\rm 2}/\bar{\kappa}_1=d^s$. 

To summarize, the mismatch between the EPL model and the deflector's true mass profile could potentially bias the lens model results via lensing degeneracies. We note that, even though we only explicitly show how the MST biases the lens modeling here for clarity; the SPT, which is a generalized version of the MST, works in a similar fashion.

\subsection{JAM Method}
\label{sec:JAM_method}
The JAM method decomposes the total mass of each galaxy into two components: a dark matter halo and a visible galaxy. The mass profile of the dark matter halo is described by the generalized NFW (gNFW) model \citep{Cappellari13}
\begin{equation}
    \rho_{\rm DM}(r)=\rho_{\mathrm{s}} \left(\frac{r}{R_{\mathrm{s}}}\right)^{-\gamma}
        \left(\frac{1}{2}+\frac{1}{2}\frac{r}{R_{\mathrm{s}}}\right)^{\gamma-3} ,
    \label{eq:16_gnfw}    
\end{equation}
where $\rho_{\mathrm{s}}$ is the characteristic density, $r_{\mathrm{s}}$ is the scale radius and $\gamma$ controls the central density slope of the dark matter halo. For $\gamma=1$, the gNFW model reduces to the well known NFW profile \citep{nfw_profile}. For the visible galaxy, the JAM method uses the Multiple Gaussian Expansion \citep[MGE]{Cappellari02} to fit the SDSS r-band image where
\begin{equation}
    \Sigma_r(x',y') = \sum_{k=1}^N
    {\frac{L_k}{2\pi\sigma^2_k q_k'} \exp
    \left[
        -\frac{1}{2\sigma^2_k}
        \left(x'^2 + \frac{y'^2}{q'^2_k} \right)
    \right]} ,
    \label{eq:17_mge}    
\end{equation}
where $L_k$ is the total luminosity of the $k$th Gaussian component.  $\sigma_k$ and $q_k$ are the dispersion and projected axis ratio of the $k$th Gaussian component and $N$ the total number of Gaussian components used in fitting. The brightness distribution of galaxies given by MGE fitting can be transformed into the mass distribution of galaxy components by assuming a constant mass to light ratio.

Given a set of model parameters (e.g.\ gNFW parameters, MGE parameters, mass to light ratio, etc.) one can derive the total mass distribution of galaxies and predict the second moment of the velocity distribution that is observed via the Jeans Anisotropic model \citep{Cappellari08}. Using the ensemble MCMC sampler emcee \citep{Foreman-Mackey13} one can find the model solution which best fits the observed second moment map deduced from the MaNGA IFU data \citep{Hongyu18}.

\subsection{Mock Lens Generation}
\label{sec:gen_data}
We generate mock lenses using the surface mass density maps of ETGs in the SDSS-MaNGA project, which are derived by \citet{Ran19} with the JAM method. The redshifts of ETGs in MaNGA are from $\sim0.02$ to $\sim0.1$ (median value $\sim0.06$), whereas the typical redshift of the SLACS lens galaxies (which are also ETGs) is $\sim$0.2. If the internal structure of ETGs has a dramatic evolution from redshift $\sim$0.2 to $\sim$0.02, using MaNGA ETGs to represent SLACS lenses would be unreasonable. Fortunately, results from numerical simulations \citep{Wang19} and observations \citep{slope_Koopmans06,slope_Koopmans09,slope_ruff11,slope_Sonnenfeld13,slope_lirui18,slope_chen19} indicate that no obvious redshift evolution exists from $z=1$ to $z=0$ and we therefore anticipate that our mock lenses will be representative of SLACS lenses. To ensure our mock lenses are most similar to those in the SLACS sample, we ``move'' a MaNGA ETG to redshift 0.2 and suppose that it acts as a strong lens with a background source at redshift 0.6. We require that the final MaNGA ETGs used to generate the mock lenses meet the following criteria:
\begin{enumerate}
\item{The Einstein radius of the mock lenses ranges between 0.6$\arcsec$ to 2.0$\arcsec$.}
\item{\reply{The luminosity-weighted velocity dispersion within the half-light radius of each ETG} is larger than $150$ \,km\,s$^{-1}$.} % Can you quantify this better?
\item{The Sersic index \citep{sersic1963} is larger than 3 to ensure the galaxy is an ETG.}
\item{The results of JAM modeling are labeled as Grade-A in \citet{Hongyu18}, to ensure that the dynamical reconstruction is reliable}.
\end{enumerate}
We use the surface mass density map ($\Sigma$) used in \citet{Ran19} divided by the critical density of the lens system ($\Sigma_{\mathrm{crit}}$) to compute the convergence
\begin{equation}
    \kappa = \frac{\Sigma}{\Sigma_{\mathrm{crit}}}
    \label{eq:18_def_kappa},    
\end{equation}
where $\Sigma_{\mathrm{crit}}$ is given by
\begin{equation}
    \Sigma_{\mathrm{crit}} = \frac{c^2}{4 \pi G} \frac{D_{\mathrm{s}}}{D_{\mathrm{l}} D_{\mathrm{ls}}} .
    \label{eq:19_crit_density}    
\end{equation}
We then apply a fast Fourier transform algorithm to $\kappa$ to calculate the deflection angle map using equation~\eqref{eq:11_int_alpha}. To reduce the numerical errors of the deflection angle calculation, we apply the following scheme:
\begin{enumerate}
\item{\reply{The deflection angle is calculated numerically on an oversampled grid of resolution $2\times2$, where the average is taken to give the deflection angle on the coarser native image grid (0.05\arcsec/pixel).}}
\item{To avoid the boundary effect shown in \citet{boundary_shear_effect}, we perform the Fourier transform on a grid of size $40^{\prime\prime} \times 40^{\prime\prime}$ \reply{whose half-width} is at least $10$ times larger than the typical Einstein radius\footnote{We demonstrate the necessity of taking a larger convergence map size to robustly estimate the deflection angle in Appendix~\ref{sec:appdix_defl}.}}.
\item{When equation~\eqref{eq:11_int_alpha} is used for numerical convolution operation, the size of the kernel is twice the size of the deflection angle map.}
\end{enumerate}
The deflection angle map derived from the above method is used to generate mock lenses. The ideal lensed image, $I_{\rm ideal}(\boldsymbol{\theta})$, can be expressed as
\begin{equation}
    I_{\rm ideal}(\boldsymbol{\theta}) = I_{\mathrm{s}}(\boldsymbol{\beta}) = 
    I_{\mathrm{s}}(\boldsymbol{\theta}-\boldsymbol{\alpha}) \,,
    \label{eq:20_form_lensed_image}    
\end{equation}
where $I_{s}(\boldsymbol{\beta})$ is the brightness distribution of the source galaxy in the source-plane. In this work, we use a {\it single} Sersic component to represent $I_s$. The expression of a Sersic profile is
\begin{equation}
    I(R) = I_{\mathrm{e}} ~ \text{exp} \left\{ -b_{\rm n} \left[\left(\frac{R}{r_{\mathrm{e}}}\right)^{1/n} - 1\right] \right\} ,
    \label{eq:21_sersic}    
\end{equation}
where $r_{\mathrm{e}}$ is the half-light radius, $I_{\mathrm{e}}$ is the brightness at $r_{\mathrm{e}}$, $n$ is the Sersic index and $b_{\rm n}$ is a coefficient that only depends on $n$. We can introduce ellipticity into equation~\eqref{eq:21_sersic} via the coordinate transformation described by equation~\eqref{eq:5_ellip_trans}. To create our 50 mock lenses, we always assume $r_{\mathrm{e}} = 0.15^{\prime\prime}$, $n=1.0$, and axis ratio $q=0.7$ \citep{Newton11}. The position $(x_{\mathrm{s}}, y_{\mathrm{s}})$ of the source galaxy is drawn from a Gaussian distribution with mean $0^{\prime\prime}$ and a standard deviation $0.1^{\prime\prime}$, and the position angle is uniformly selected between $0^{\circ}$ and $180^{\circ}$. By using this set of parameters, the final lensed images we generate usually have extended arc structures. The constraining power provided by lensing is related to the geometry configuration of the lens system (cusp, fold, two images, Einstein cross) and the extended arc can provide more information on a deflector's mass distribution, which helps to reduce systematic errors in the final modeling result \citep{Tagore18}. Since this work focuses on evaluating the systematic errors of galaxy-galaxy lens modeling under the EPL assumption, we choose to model a variety of lensing configurations and do not specifically assess the influence of a given lensing configuration on our modeling results.

\reply{To simulate observational effects, the image pixel size is set to 0.05\arcsec (HST-quality). The ideal lensed image derived by equation~\eqref{eq:20_form_lensed_image} is convolved with a Gaussian Point Spread Function (PSF, standard deviation: $0.05^{\prime\prime}$).} We assume a sky background of 0.1 $\rm counts\,s^{\rm -1}$ and an exposure time of $\rm 840\,s$. Background noise from the skylight and Poisson noise from the target source is also added to mock images. We manually adjust the brightness $I_\mathrm{e}$ of the source galaxy to ensure that the peak signal-to-noise ratio of each lens in our final sample is $\sim$50, which roughly corresponds to the highest signal-to-noise observations in the SLACS project \citep{Bolton_sample_08}.

To compute true time-delay values of our mock lenses at a given set of image positions, we need true lens potential values $\psi$. We use equation~\eqref{eq:12_int_psi} to calculate the lens potential map from the $\kappa$ map and apply the strategy described previously to reduce the numerical error on the Fast Fourier transform. The lens potential at any position on the lens plane can then be obtained via interpolation.

\begin{figure*}
    \centering
	\includegraphics[width=\textwidth]{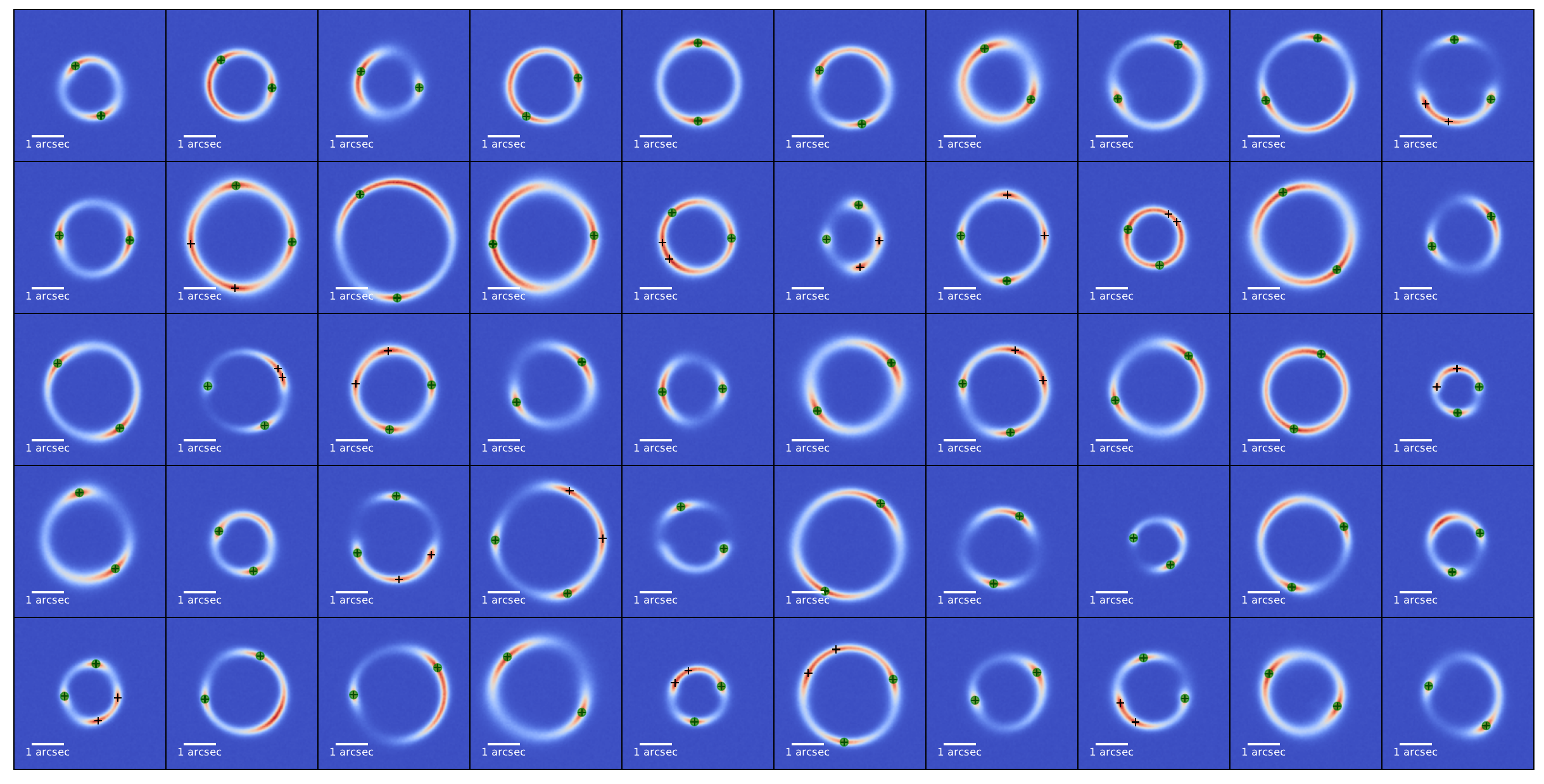}
    \caption{Images of 50 ``MaNGA lenses''. For each lens, locations marked by the black crosses represent the lensed positions of the source center; the points marked by the green circle are used to calculate the excess time-delay between different images as shown in Figure~\ref{fig:13_td}. The scale-bars mark the angular scale of 1 arcsecond.}
    \label{fig:1_mock_lens_image}
\end{figure*}

Figure~\ref{fig:1_mock_lens_image} shows the sample of 50 lenses that we have simulated, which we refer to hereafter as the ``MaNGA lenses'' for short. All our lensing images have extended arc structures, which make them a good probe of the underlying mass distribution. The green circles mark the lensed positions of the center of the source galaxies. The time-delay values between these positions are calculated and used when we test the measurement of $H_{\rm 0}$.

\begin{figure}
    \centering
	\includegraphics[width=\columnwidth]{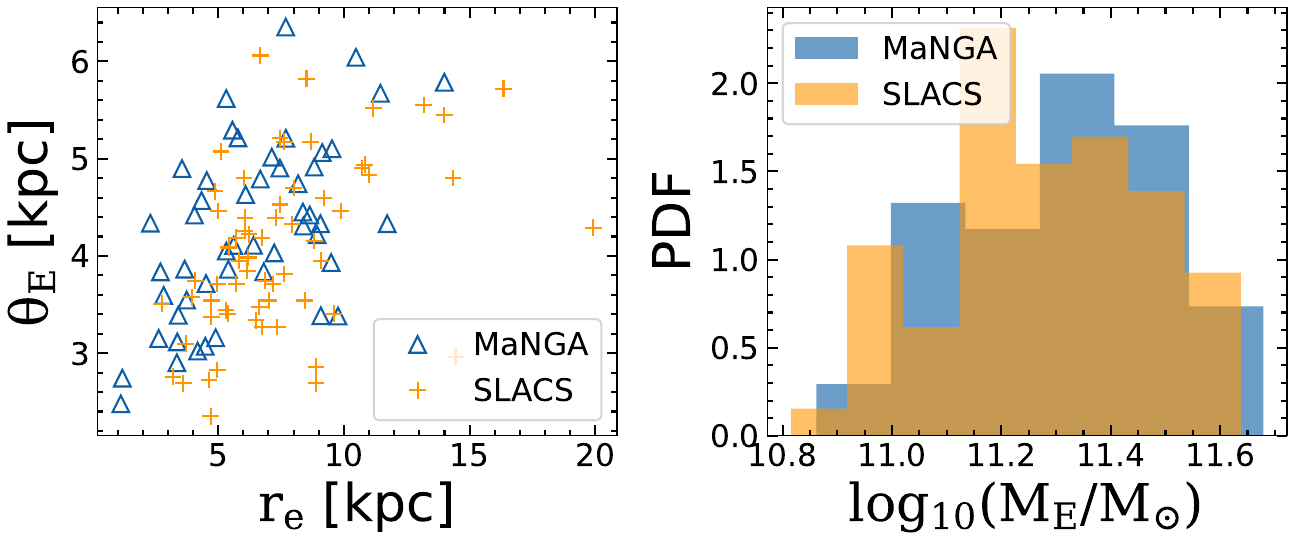}
    \caption{The left panel shows the dependence of the Einstein radius on the half-light radius of the lens galaxy. Each data point corresponds to a lens system, blue represents ``MaNGA lenses'' and orange represents SLACS lenses. On the right panel, the blue histogram shows the Einstein mass distribution of ``MaNGA lenses'' and the orange histogram represents the distribution of SLACS samples.}
    \label{fig:2_mock_reff_thetaE}
\end{figure}

Figure~\ref{fig:2_mock_reff_thetaE} shows the relationship between the Einstein radius (measured from the Equation~\eqref{eq:14_def_thetaE}) and the half-light radius of the MaNGA lenses (measured from the Sersic fit), and the probability density distribution of their Einstein mass (right panel). For comparison, we also show the values of lenses in SLACS (in orange). One can find that the distribution of properties of our mocks and those of lenses from SLACS are quite similar.

% \footnotetext{During the lens modeling, we require a feasible lens mass solution to fulfill the following condition: after tracing the images marked by the black cross back to the source plane, those points should locate within a small range (a circle with a radius of $0.3^{\prime\prime}$). This method can improve the modeling speed and avoid solutions with non-physical sources.}

\section{Modeling methodology}
\label{sec:lens_model_method}
We use {\tt PyAutoLens} \citep{pyautolens}\footnote{The {\tt PyAutoLens} software is open-source and available from \url{https://github.com/Jammy2211/PyAutoLens}} to model the simulated ``MaNGA lenses'', which is described in \citet[][N18 hereafter]{Nightingale18} and builds on the works of \cite{Warren03, Suyu06, Nightingale15}. {\tt PyAutoLens} uses an empirical Bayes framework and a technique called non-linear search chaining to compose pipelines which break the lens modeling procedure into a series of simpler model-fits. This allows a user to begin modeling a system with a simple lens model (e.g. an isothermal mass profile and a Sersic source) and via a sequence of non-linear searches gradually increase the model complexity, so as to eventually fit the desired more complex lens model (in this work, an EPL + shear using a pixelized source reconstruction). Non-linear search chaining is implemented in {\tt PyAutoLens} via the probabilistic programming language {\tt PyAutoFit}\footnote{\url{https://github.com/rhayes777/PyAutoFit}} and it provides three main advantages for lens modeling:
\begin{enumerate}

\item{For complex lens models, the non-linear search may not be able to sample parameter space sufficiently well to find the global maximum-likelihood solution and may instead infer an inaccurate local maximum. Fitting simpler lens models in early searches helps mitigate this, since the results of fitting a simple model can guide the choice of prior for the more complex models fitted by subsequent non-linear searches. By providing the non-linear search a better initialization of where to search the parameter space, one can therefore reduce the risk of inferring a local maximum. This allows us to achieve fully automated lens modeling in this work.}

\item{The settings of {\tt PyAutoLens} and the non-linear search are customized to give fast and efficient model-fits in early stages of a pipeline (where a precise quantification of the model parameters and errors is not required) and perform a more thorough analysis later on (where a precise estimate of the lens model parameters and errors is desired)}.

\item{In addition to the results of the lens model itself, other results of the model-fit are also passed to the fits performed later on. For example, we use {\tt PyAutoLens}'s `hyper-mode', which fully adapts the model to the characteristics of the imaging data which is being fitted by passing the model-image of the lensed source galaxy inferred by earlier fits to later fits.}
\end{enumerate}

We use a single Sersic profile to represent the brightness of the source galaxy when we simulate the "MaNGA lenses". We can therefore in principle ``perfectly'' represent the source's light using a Sersic source model. However, for real strong lens observations, the brightness distribution of the source is often irregular and complex, which necessitates the use of pixelized source models \citep{Nightingale18}. To ensure our results can generalize to observations of real strong lenses, we therefore use a pixelized source reconstruction to fit the lens model, even though a Sersic source would suffice. We use the adaptive brightness pixelization and regularization scheme provided by {\tt PyAutoLens} to do this. The adaptive pixelization congregates source-pixels to the regions where the source is located, effectively improving the resolution of the source reconstruction. Adaptive regularization makes it so that {\tt PyAutoLens} can reduce the regularization strength of the source in its central regions relative to its outskirts, which is important for reconstructing sources with compact or clumpy structures.

To perform model-fitting via non-linear search chaining, we use the ``SLaM'' (Source, Light, and Mass) pipelines distributed with {\tt PyAutoLens} to model our ``MaNGA lenses''. In brief, our lens modeling pipeline divides the lens modeling process into three steps:
\begin{enumerate}
\item{The mass distribution of the lens is modeled by a singular isothermal ellipsoid + shear. The brightness distribution of the source is modeled by a Sersic component. The lens modeling results of this step tell us the approximate mass distribution of the lens, which is next used to initialize the pixelized source model.}
\item{Based on the results of the mass model in step (i), the pixelized source model is fitted.}
\item{based on the optimal pixelized source model obtained in step (ii), fit the lens mass with the EPL + Shear model.}
\end{enumerate}
This pipeline therefore finishes by fitting the EPL + shear model, on which the majority of our results are based. During the lens modeling, we require a feasible lens mass solution to fulfill the following condition: after tracing positions marked by black crosses in Figure~\ref{fig:1_mock_lens_image} back to the source plane, those points should locate within a small range (a circle with a radius of $0.3^{\prime\prime}$)\footnote{\reply{Although we do not explicitly use the position of the selected lensed images to constrain the lens mass model (i.e, integrate the position constraints into the Gaussian-form likelihood function), we found that after tracing the selected lensed images shown in Figure~\ref{fig:1_mock_lens_image} (marked by black crosses) back to the source plane (using the best-fit lens model based on the extended arc), the distance between them does not exceed 0.05\arcsec. Therefore, we expect that the extended arc has already efficiently constrained the lens mass model, and the positions of the counter images will not provide too much extra contribution.}}.
\begin{figure*}
    \centering
	\includegraphics[width=\textwidth]{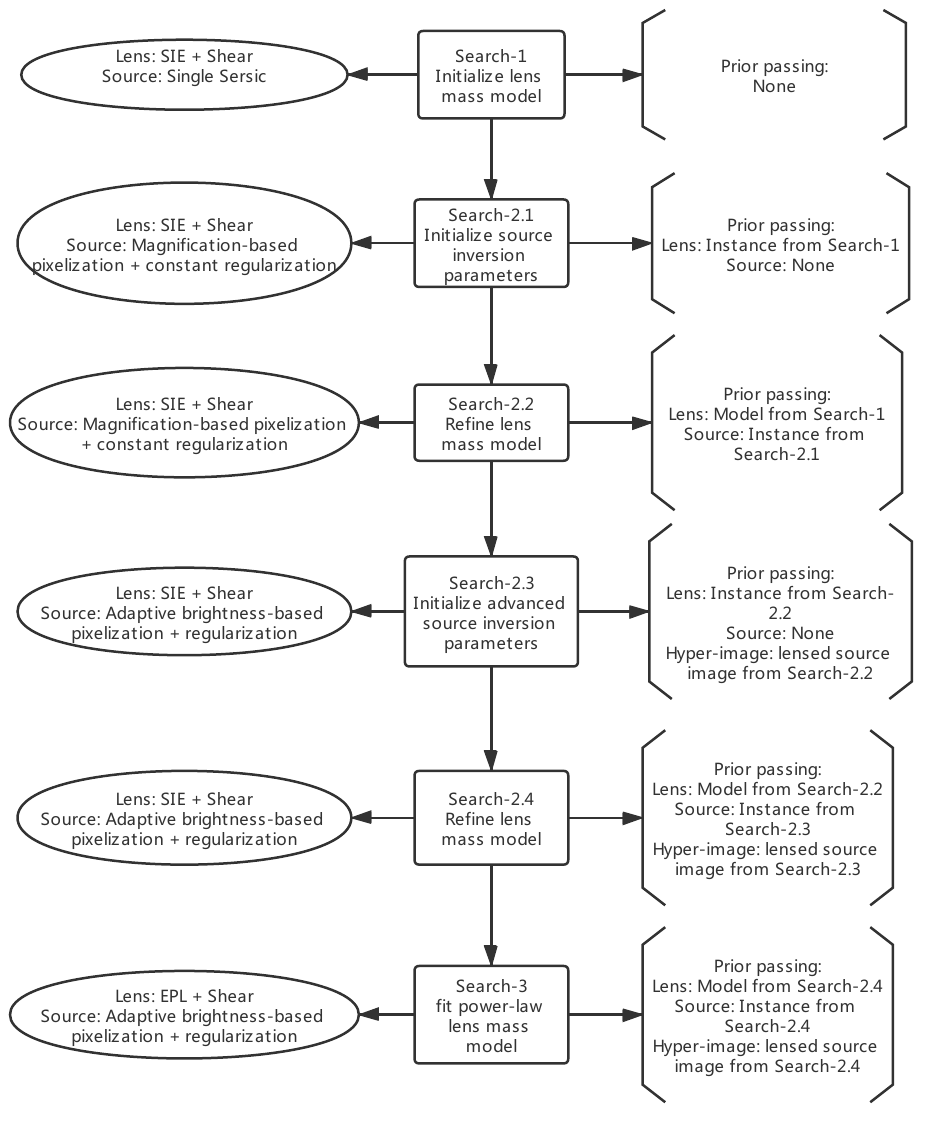}
    \caption{This figure shows the modeling process using the advanced pixelized source model provided by {\tt PyAutoLens}. The middle box shows the purpose of the current non-linear search. The model component of the current non-linear search is recorded in the left ellipse. The bracket on the right shows the prior transfer of the current non-linear search, where ``instance'' represents that the parameters of the model inherit the best-fit results of previous non-linear searchs, and keep fixed during modeling; While ``Model'' is similar to ``instance'', but inherits parameters in the Gaussian prior way.}
    \label{fig:3_lens_pipe}
\end{figure*}
This trick can improve the modeling speed and avoid solutions with non-physical sources. For a more detailed overview of the modeling procedure, we refer to the flow chart shown in Figure~\ref{fig:3_lens_pipe} \footnote{For more specific prior-passing configurations, see the online material: \url{https://github.com/Jammy2211/autolens_workspace}.}.

The goal of the paper is to study the systematics errors induced by the EPL mass model when the true mass distribution of a lens system is more complex. Since the more flexible pixelized source model can 'absorb' some of the residual signals due to the mass model mismatch, it is important to check whether this affects the lens mass parameters we estimate. Therefore we performed an additional set of model fits whose source light is represented by a parametric Sersic. The lens mass parameters determined by model fits with the parametric source model are statistically consistent with those derived by assuming the pixelized source model, hence we only report the results of the pixelized source model.

\section{Results and discussion}
\label{sec:results_and_discusion}
This section presents the lens modeling results of our mock ``MaNGA lenses'' under the assumption of the EPL + shear lens model. Section~\ref{sec:mass} lists the reconstruction results of the lens mass model, including the measurement of Einstein radius (Section~\ref{sec:mass_thetaE}), shear (Section~\ref{sec:mass_shear}), and density slope (Section~\ref{sec:mass_slope}). Section~\ref{sec:source} shows how the source reconstruction is influenced by the EPL assumption. We show the anatomy of individual lens systems for a typical case and an outlier in Sections~\ref{sec:source_typical_case} and \ref{sec:source_outliers}, respectively. The statistical results of source reconstructions are presented in Section~\ref{sec:source_stat}. We will discuss how the EPL mass assumption affects the estimation of time-delays and $H_0$ in Section~\ref{sec:time_delay}.

%----------------------------------------------------------mass model
\subsection{Mass model reconstruction}
\label{sec:mass}
In this section, we discuss the systematic errors caused by assuming the EPL + shear mass model for measurements of the Einstein radius (Section~\ref{sec:mass_thetaE}), shear (Section~\ref{sec:mass_shear}), and inner density slope (Section~\ref{sec:mass_slope}).

\subsubsection{Einstein radius}
\label{sec:mass_thetaE}
\begin{figure}
	\includegraphics[width=\columnwidth]{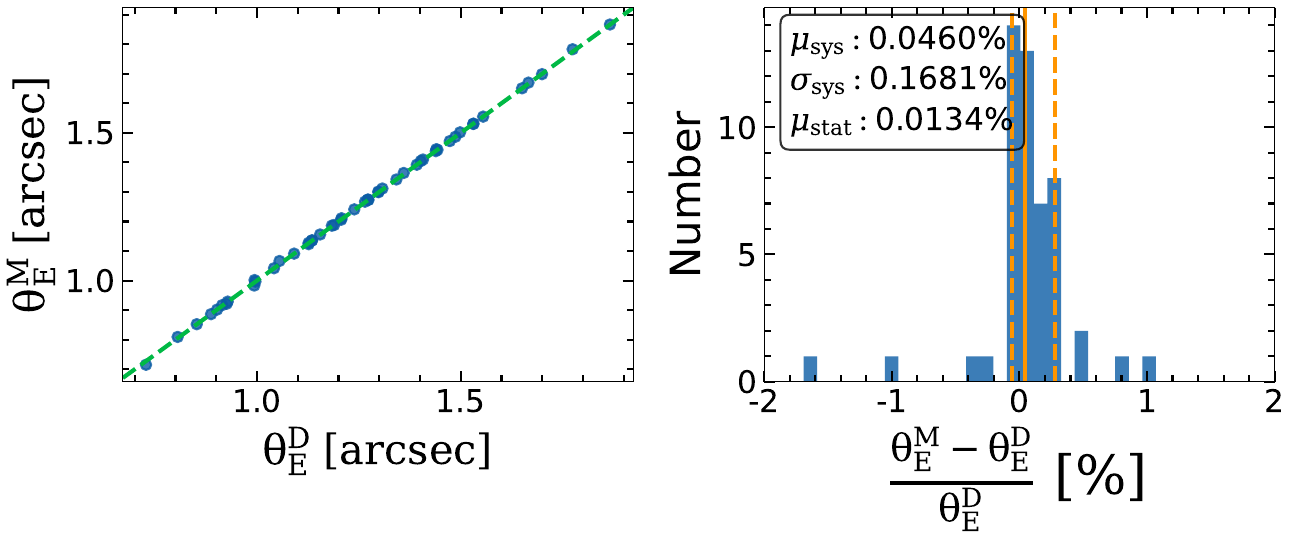}
    \caption{The left panel compares the model value of Einstein radii ($\rm \theta_{\rm E}^M$) with the ground truth ones ($\rm \theta_{\rm E}^D$, defined by Equation~\eqref{eq:14_def_thetaE}). Each data point represents the model reconstruction of a lens system; a data point lies on the green dashed line if the model value matches the true value perfectly. Since the statistical error on the Einstein radius is small (only of order 0.01 \%), we don't include the error bar for each data point on the left panel, and only present the median of the relative statistical error ($\rm \mu_{stat}$) in the legend of the right panel. The right panel shows the number distribution of relative systematic error. The orange vertical lines show the median value of the relative systematic error distribution, and the two dot-orange vertical lines mark the 68\% confidence interval ([16\%, 84\%] percentile). We show the median ($\rm \mu_{sys}$) and dispersion ($\rm \sigma_{sys}$) (i.e, the standard deviation) of the systematic error in the legend.}
    \label{fig:9_lens_thetaE_recon}
\end{figure}

Figure~\ref{fig:9_lens_thetaE_recon} shows the inferred Einstein radius of our 50 ``MaNGA lenses'' under the assumption of EPL + shear mass model. We find that the relative systematic error of the model is only $0.05\% \pm 0.17\%$ (68\% confidence level), indicating that the estimation of Einstein radius is unbiased. Given that the magnitude of the statistical error is only of order 0.01\% and its error bar is almost invisible, we only show the typical value of the statistical error in the legend of the right panel. Summing both statistical and systematic errors in quadrature, our results indicate that lensing can determine the Einstein radius with an accuracy of $\sim0.1\%$.

As a demonstration, we consider a constraint on the post-Newtonian gravity parameter $\gamma_\mathrm{ppn}$, following the method of \cite{ppn_bolton06,ppn_Schwab_2010, ppn_caoshuo17, ppn_Collett2018, ppn_yang2020}. This compares the mass of galaxies from lensing and stellar dynamics 
\begin{equation}
M_\mathrm{dyn}=\frac{1+\gamma_\mathrm{ppn}}{2}~M_\mathrm{lensing}^\mathrm{GR}.
\label{eq:ppn}
\end{equation}
Assuming that the measurements of stellar dynamics are error-free, such that the only source of bias is from lensing, \reply{our $\sim0.1\%$ relative error on Einstein radius translates into an error of $\sim0.004$ on $\gamma_\mathrm{ppn}$ (see Appendix~\ref{sec:ppn_err})}. This is an order of magnitude smaller than current measurement uncertainty \citep[$\gamma_\mathrm{ppn}=0.97\pm0.09$,][]{ppn_Collett2018}. Therefore, use of the EPL mass model does not currently limit this test of gravity.

\subsubsection{Shear}
\label{sec:mass_shear}
\begin{figure}
	\includegraphics[width=\columnwidth]{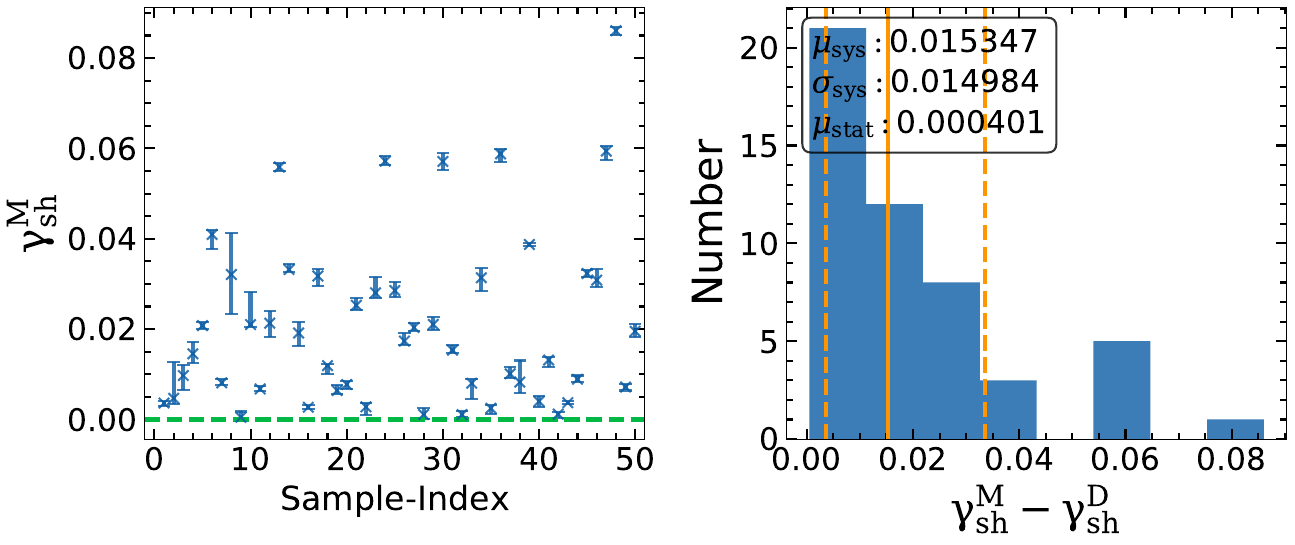}
    \caption{Similar to Figure~\ref{fig:9_lens_thetaE_recon}. The shear strength reconstructed by the lens model is compared with the true value. The error bars represent the $3\sigma$ limit of each lens. Given that we don't include the external shear component when simulating mock data, the ground truth value of the shear strength is zero ($\rm \gamma_\mathrm{sh}^\mathrm{D}=0$), which is marked by the green dashed line. The right panel shows the distribution of the absolute systematic error.}
    \label{fig:10_shear_recon}
\end{figure}
\begin{figure}
	\includegraphics[width=\columnwidth]{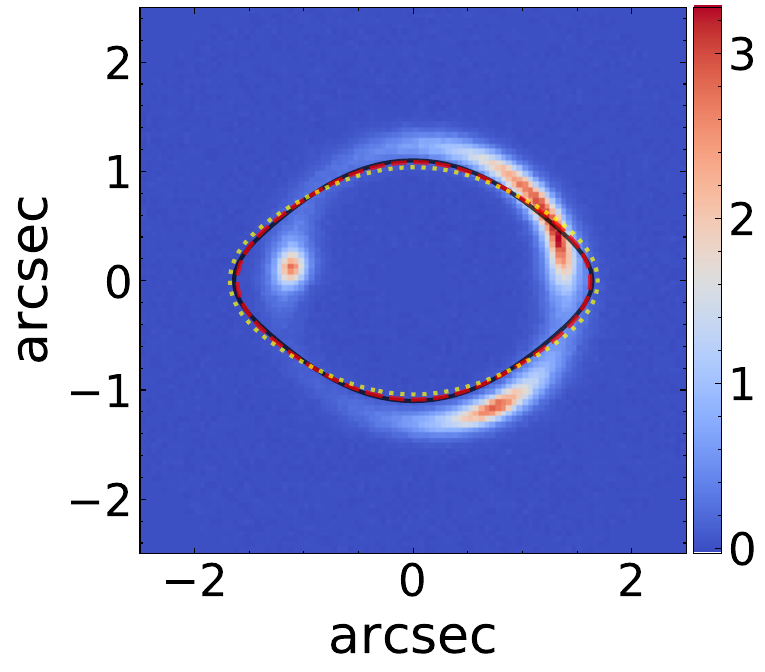}
    \caption{The best-fit lens model of system ``3024\_9039-1902''. The black line represents the true critical line of the mock data. The yellow dotted line marks the critical line given only by the EPL model, while the red dashed line is the critical line predicted by the EPL+shear model.}
    \label{fig:11_shear_indivi_check}
\end{figure}
The distribution of shear strength measured for each lens model is shown in Figure~\ref{fig:10_shear_recon}. Although we did not add an external shear term when generating mock data, We have to include a shear component when doing lens modeling otherwise we get a bad image fit with significant residuals. The median inferred shear magnitude for our 50 ``MaNGA lenses'' is 0.0153 and the standard deviation is 0.0150. Recall that when we simulated the mock lenses, the surface mass distribution of their stellar components was given by the MGE model which did not possess homoeoidal elliptical symmetry \footnote{Each elliptical Gauss component in MGE shares the common center and position angle, but the axis-ratio is allowed to be different.}. For the dark matter the gNFW model was used, which is circularly symmetric. The total surface mass density of simulated lenses therefore does not satisfy elliptical-symmetry and the EPL lens model by itself is unable to describe this departure from elliptical symmetry. However, the shear term can mimic the effect of deviating from the elliptical-symmetry to the first-order approximation \citep{Keeton1997}. This is illustrated more clearly in Figure~\ref{fig:11_shear_indivi_check}. For the mock lens shown in Figure~\ref{fig:11_shear_indivi_check}, the critical line of the MGE+gNFW simulated mass distribution has a `boxy' shape (black line) because it does not satisfy elliptical symmetry. The EPL model by itself gives a critical line (yellow dotted line) with an elliptical shape that closely traces the black line, but is unable to trace it perfectly. By including a shear, the EPL+shear model's critical line (red dashed line) more closely approximates the true profile's `boxy' shape. Therefore, the shear detected in this work is an internal shear caused by the $\kappa$ field of the lens system itself that does not satisfy the elliptical-symmetry, rather than an external shear caused by line of sight structures.

Cosmic shear can be used to infer the mass perturbations along the line of sight, making it an important probe of the large-scale structure of the universe. \citet{shear_birrer17,shear_birrer18} proposed using strong gravitational lens systems with extended ring structures to measure cosmic shear, where a combination of this measurement with weak lensing can illuminate the different systematic errors between the two methods. Our results here show that when using strong gravitational lenses to measure the cosmic shear, it may be important to ensure the method is not subject to systematic errors induced by the lens not possessing elliptical-symmetry.

%--------------------density slope section
\subsubsection{Density slope}
\label{sec:mass_slope}

\begin{figure}
	\includegraphics[width=\columnwidth]{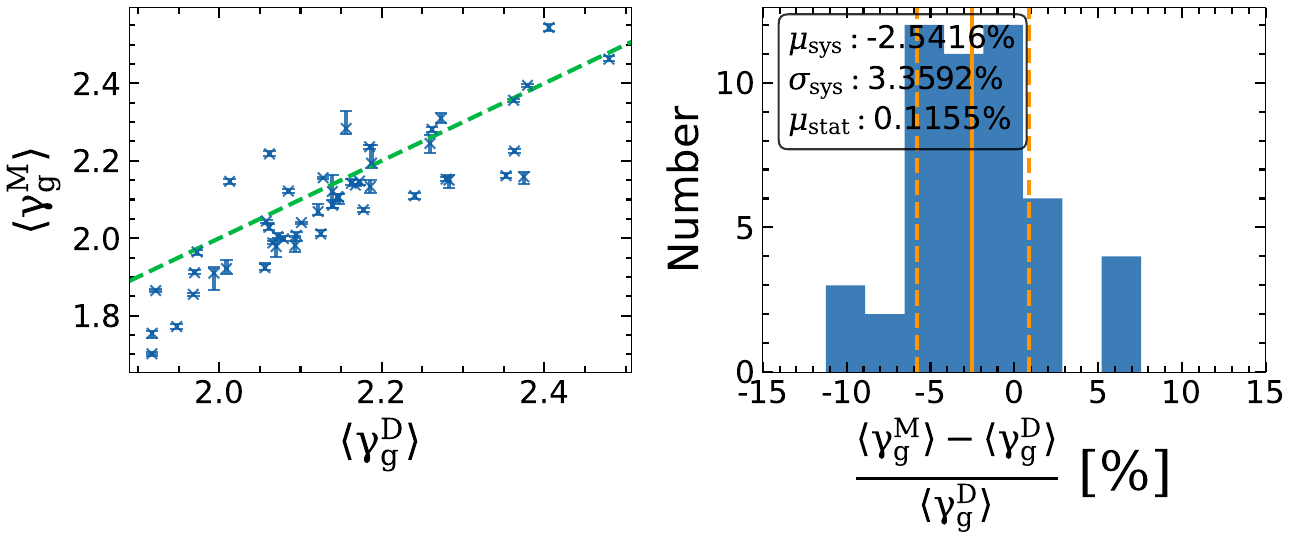}
	\includegraphics[width=\columnwidth]{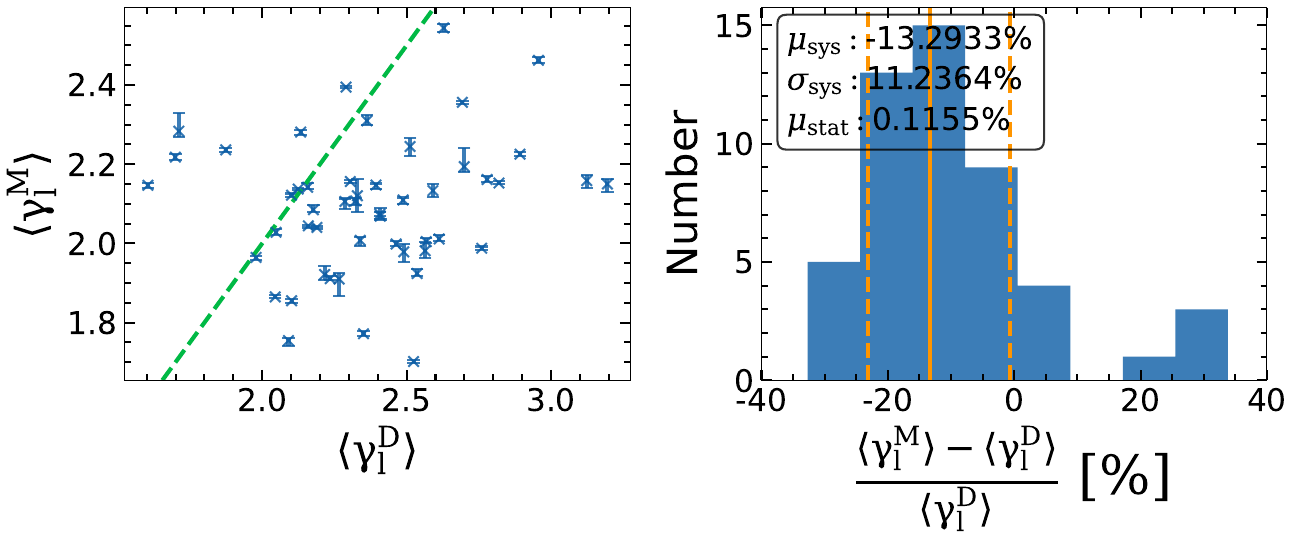}
    \caption{
    \textit{Top panel:} Reconstruction of the deflector's ``global density slope'' $\rm \langle\gamma_g\rangle$ within the Einstein radius. Error bars in the left panel show $3\sigma$ limits.
    \textit{Bottom panel:} Reconstruction of the deflector's local density slope $\rm \langle\gamma_l\rangle$ near the Einstein radius.
    Green dashed lines indicate the 1:1 relation.
    }
    \label{fig:slope_recon}
\end{figure}

Our lens models accurately recover the global density slope, defined as the mass-weighted density slope within the Einstein radius (equation~\eqref{eq:15_def_slope}). Averaging across our mock sample, the inferred global density slope is smaller than the input value by only $2.5\%$, with a root-mean-square scatter of $3.4\%$ (top panel of Figure~\ref{fig:slope_recon}).
However, the local density slope at the Einstein radius, defined as the mass-weighted density slope within the radial interval [$0.8\times\theta_{\rm E}$, $1.2\times\theta_{\rm E}$] (equation~\eqref{eq:15_def_slope_2}) is less accurately recovered. It is underestimated (on average) by $13.3\%$ with a scatter of $11.2\%$ (bottom panel of Figure~\ref{fig:slope_recon}).

\begin{figure}
	\includegraphics[width=\textwidth]{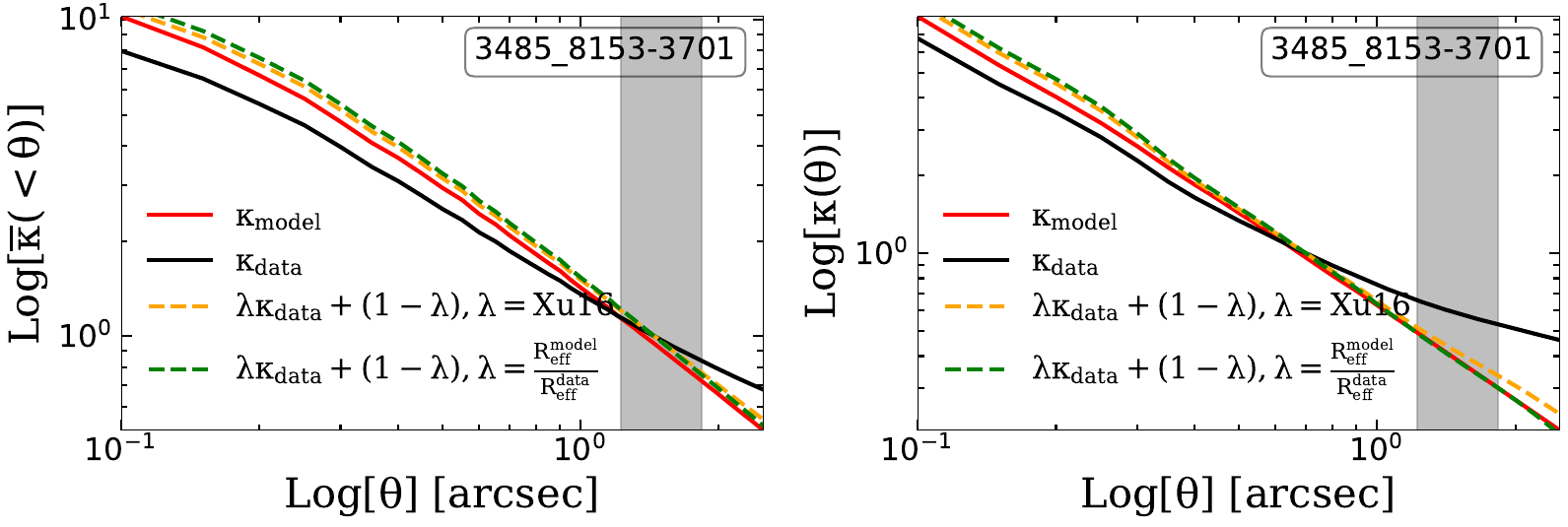}
    \caption{The convergence profile of the lens system ``3485\_8153-3701'', whose lens model bias due to the oversimplified EPL mass model can be understood via the MST. Left: the red solid line represents the radial profile of the average convergence within certain radii ($\bar{\kappa}(<\theta)$) predicted by the lens modeling. The intrinsic average convergence profile of data is shown with the black solid line. Other lines are mass-sheet transformed solutions of the black line. The corresponding $\lambda$ factor of MST for the orange dashed line \reply{($\lambda=1.42$)} is given by Equation~\ref{eq:mst_lambda}, which makes its deflection angle field most similar to the power-law functional form in the vicinity of the Einstein ring marked by the shaded grey. The $\lambda$ factor for the green dashed line \reply{($\lambda=1.49$)} is the ratio of the model's source size to the input true value. Right: the (non-averaged) radial density profile $\kappa$, lines with different styles one-to-one correspond to the right panel.}
    \label{fig:7_outliers_avg_kappa_prof}
\end{figure}

It may seem counter-intuitive that we measure the density slope better for the whole lens than at the specific radius of the source flux, similar results have also been observed in \cite{Riordan2020}. Following \cite{Kochanek20}, we had expected a measurement of global density slope to be just an extrapolation of the local mass distribution inferred by lensing with the EPL functional form. Our result is explained by the 
%The poorly constrained local density slope can be understood. Since there is a mass distribution 
mismatch between the ideal EPL model and real lenses. Pure image-based lens modeling under the EPL assumption will infer the deflector's mass profile as approximately the true mass profile after an SPT transform, because this transformed mass profile more closely follows the functional form of the EPL model in terms of the deflection angle field (or approximately average convergence within radius $\rm \theta$, $\rm \bar{\kappa}(<\theta)$, see Equation~\eqref{eq:4_alpha_kappa_mean}) near the Einstein ring.

\reply{This mismatch directly leads to the 11 ``outliers'' in our modeling results, whose physical parameters such as the source size and the lens density slope are significantly misestimated.} Furthermore, we find the modeling results of 9 outliers can be approximately understood via the classical MST, a special case of the SPT.
As an example, we present the anatomy of a typical outlier (``3485\_8153-3701'') in Figure~\ref{fig:7_outliers_avg_kappa_prof}, whose local density slope is significantly misestimated. The left panel of Figure~\ref{fig:7_outliers_avg_kappa_prof} shows the radial density profile $\bar{\kappa}(<\theta)$ of the deflector. The black solid line shows the deflector's true mass profile while the red solid line represents that inferred by the best-fit EPL + shear lens model. Clearly, the reconstructed density profile mismatches the input one although the overall fitting of the image is good. For this system, the mass distribution mismatch between the ideal EPL model and ``MaNGA mock lenses'' can be compensated by a mass-sheet transform of $k'=\lambda\kappa + (1-\lambda)$. We apply the MST to the deflector's true density profile (black line) and show the resulting profiles using orange and green lines, where the $\lambda$ for the orange line is derived using the Equation~\eqref{eq:mst_lambda} and the $\lambda$ for the green line is $\lambda={R_{\rm eff}^{\rm model}}/{R_{\rm eff}^{\rm data}}$ where $R^\mathrm{model}_\mathrm{eff}$ and $R^\mathrm{data}_\mathrm{eff}$ are the source size of the model and the ground truth value, respectively (see more discussion in Section~\ref{sec:source}). One finds that both lines are nearly coincident with the one inferred by the lens model (red line) near the Einstein ring (marked by the shaded grey), and by definition, the mass profile of both green and orange lines can produce the same lensed image as that of the input mass distribution. Therefore, the mass distribution mismatch between the EPL model and our ``MaNGA mock lenses'', drives our pure image-based lens model pipeline to approach a biased solution, i.e, an ``MST solution'' of the deflector's true mass distribution. To demonstrate how the above effect directly affects the inferred local density slope, we show the $\rm \kappa$ profile of this lens system in the right panel of Figure~\ref{fig:7_outliers_avg_kappa_prof}. The local density slope of the deflector (the black solid line) near the Einstein ring is substantially different from the value inferred by the lens model (the red solid line).

\begin{figure}
	\includegraphics[width=\columnwidth]{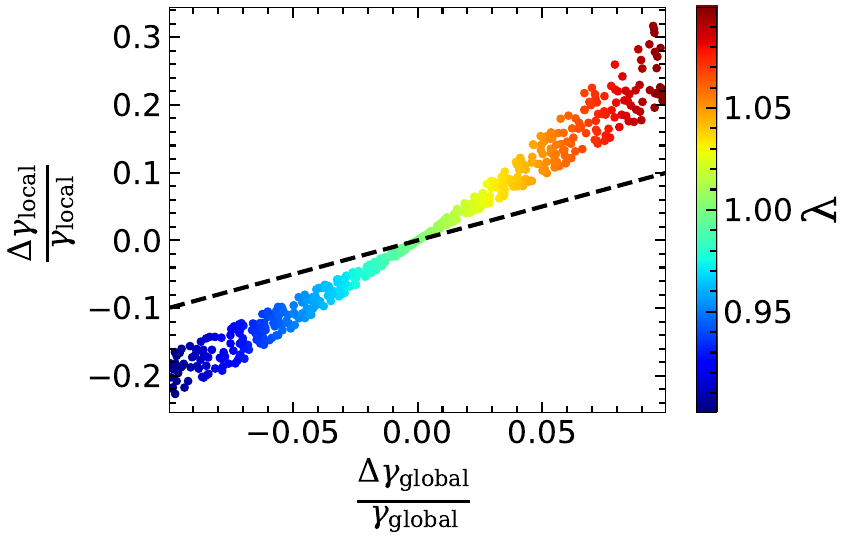}
    \caption{Each data point corresponds to an ideal spherical power-law lens, with the Einstein radius set to 1.0\arcsec, and density slope ranging from 1.9 to 2.3. A mass-sheet transformation with a $\rm \lambda$ value in the range from 0.9 to 1.1 (shown with the colour bar) is exerted on each lens. We compare the relative change of the deflector's local/global density slope before and after doing the MST. The black dashed line marks the relation in which the relative change of local density slope is equal to the global one.}
    \label{fig:local_global_slope_mst}
\end{figure}

The measurement of the global density slope is influenced by a similar effect. However, the global density slope is less impacted by the MST (or SPT). In other words, when we use an MST to rescale the deflector's true mass profile with Equation~\eqref{eq:1_MST}, the relative change of the local density slope is larger than the global density slope. We demonstrate this point via a toy model in Figure~\ref{fig:local_global_slope_mst}. We find the relative change of local density slope reaches beyond 30\% in some cases, but the relative change of global density slope is still below 10\%.

Overall, the global density slope recovered by the EPL model is sufficiently accurate for current tests of galaxy evolution \citep[Etherington et al.\ in prep.]{Wang19}. Accurately measuring local density slope, however, may require a form of mass distribution that better represents the target lens (to avoid the mass distribution mismatch discussed above), or combining with additional observables to provide extra constraints that break the degeneracy \citep{slope_Koopmans06,slope_Koopmans09,pure_lensing_Shajib20,next_gen_td2020}.

%\reply{In principle, the stellar dynamics directly constrain the 3-dimensional mass distribution, while the lensing directly measures the projected 2-dimensional mass distribution. To form a comparison between those two independent mass estimators, people may either project the 3-dimensional mass distribution given by stellar dynamics to 2-dimension or deproject the 2-dimensional  mass distribution inferred by lensing back to 3-dimension. Here, We choose the latter way, i.e, deproject the 2-dimensional density slope inferred by lensing back to the 3-dimension directly (by plus 1) and compare it to that given by stellar dynamics. Comparing the density slope in 3-dimension is more common in many previous works \citep{lmass_he20,pure_lensing_Shajib20,AndrewNewman2013,Dye08}.}

\reply{Note that stellar dynamics constrains the 3-dimensional mass distribution, while lensing measures the projected 2-dimensional mass distribution. To compare these two independent mass estimators, one can project the 3-dimensional mass distribution given by stellar dynamics to get the projected (2-dimensional) mass distribution or deproject the 2-dimensional  mass distribution inferred by lensing to infer the 3-dimensionsional distribution. Here, we choose the latter, converting the 2-dimensional density slope inferred by lensing to the 3-dimensional density slope -- which in the case of a power-law density profile means adding one to the power-law exponent. We report our findings in terms of the 3-dimensional density slope for consistency with previous work \citep{lmass_he20,pure_lensing_Shajib20,AndrewNewman2013,Dye08}.}

%-----------------------------------------------source model
\subsection{Source morphology}
\label{sec:source}
Gravitational lenses can be used as cosmic telescopes to study the structure of their highly magnified source galaxies. An important question is whether the lens modeling process reconstructs the source's morphology accurately and with high fidelity? Approaches such as the pixelized source reconstructions used in this work can reliably recover the structure of the source when the lens mass model is correct \citep{Warren03,Suyu06,Tagore14,Nightingale15,Nightingale18}. However, for real lenses the mass model will not perfectly describe the lens's true underlying mass distribution and this mismatch could potentially bias the source reconstruction, in particular via the way of the SPT. In this section, we use our ``MaNGA lenses'' test suite to evaluate systematic errors of the source reconstruction caused by the oversimplified EPL mass model.

\begin{figure*}
	\includegraphics[width=\textwidth]{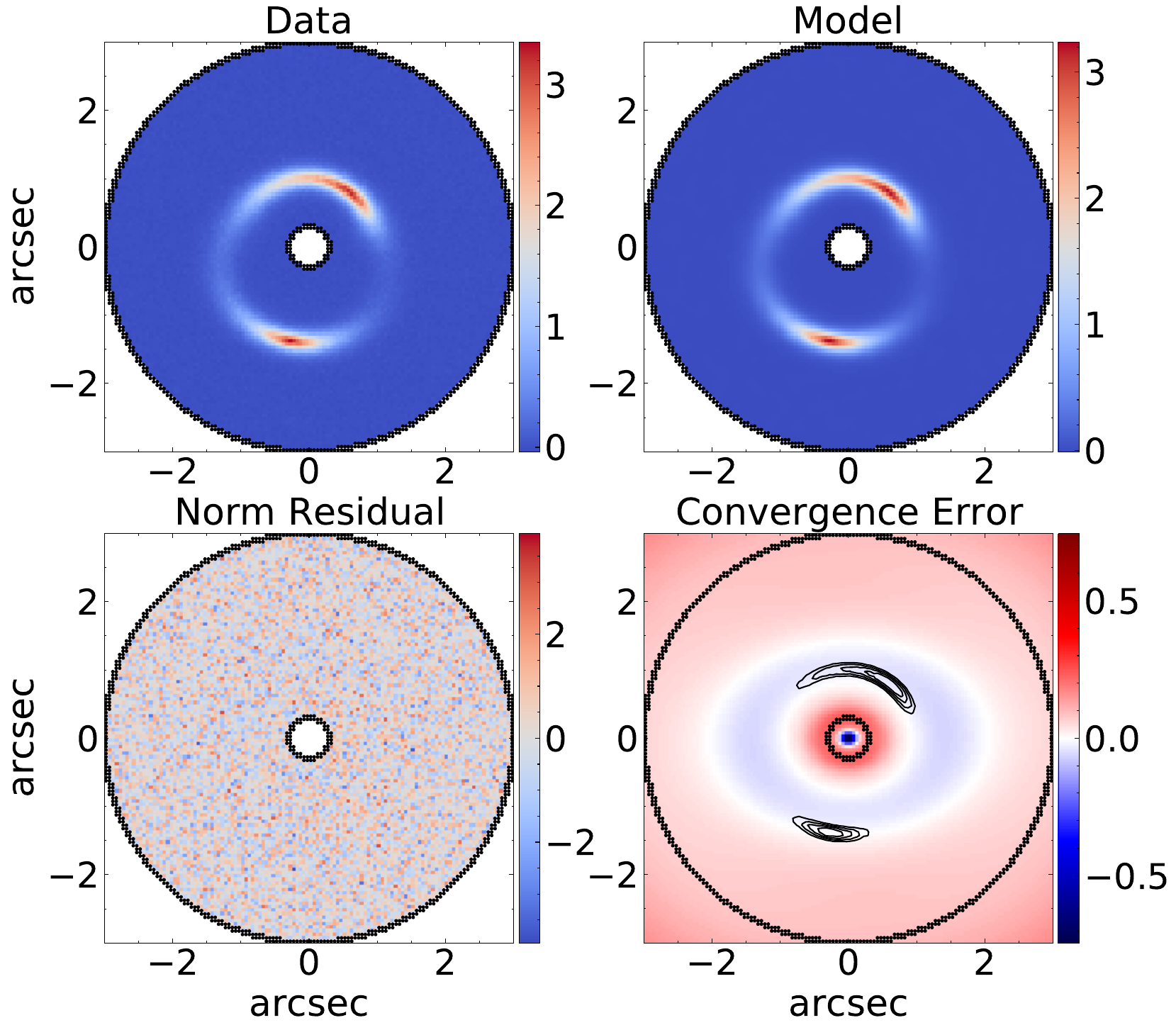}
    \caption{The lens modeling result of a typical system ``1344\_8263-9102'', which is well fit by an EPL + shear mass model. Top-left: the image of mock data $I_\mathrm{data}$; Top-right: the reconstruction of lens modeling $I_\mathrm{model}$. Bottom-left: the normalized residual map of the best-fit model, defined as $(I_\mathrm{data}-I_\mathrm{model})/I_\mathrm{noise}$, where $I_\mathrm{noise}$ is the noise map of the mock data; Bottom-right: the relative error map between the convergence map of the model ($\kappa_\mathrm{model}$) and the ground truth value ($\kappa_\mathrm{data}$), which is defined as $\kappa_\mathrm{data}/\kappa_\mathrm{model}-1$ \reply{(in accordance with \citet{Enzi20})}. Contours mark the lensed image of the mock data.}
    \label{fig:4_typical_case_model_result}
\end{figure*}

\subsubsection{Typical case: ``1344\_8263-9102''}
\label{sec:source_typical_case}
Our lens models capture the overall structure of almost all lens systems well. The fit to a typical lens-plane image (using an EPL + shear mass model and pixelized source) is shown in Figure~\ref{fig:4_typical_case_model_result}. The normalized residuals for this fit, which are shown in the bottom left panel, are almost within the $<3\sigma$ limit, indicating a good lens model. Furthermore, the inferred convergence broadly agrees with the input values in regions near strongly lensed images, as shown in the bottom right panel.

\begin{figure*}
	\includegraphics[width=\textwidth]{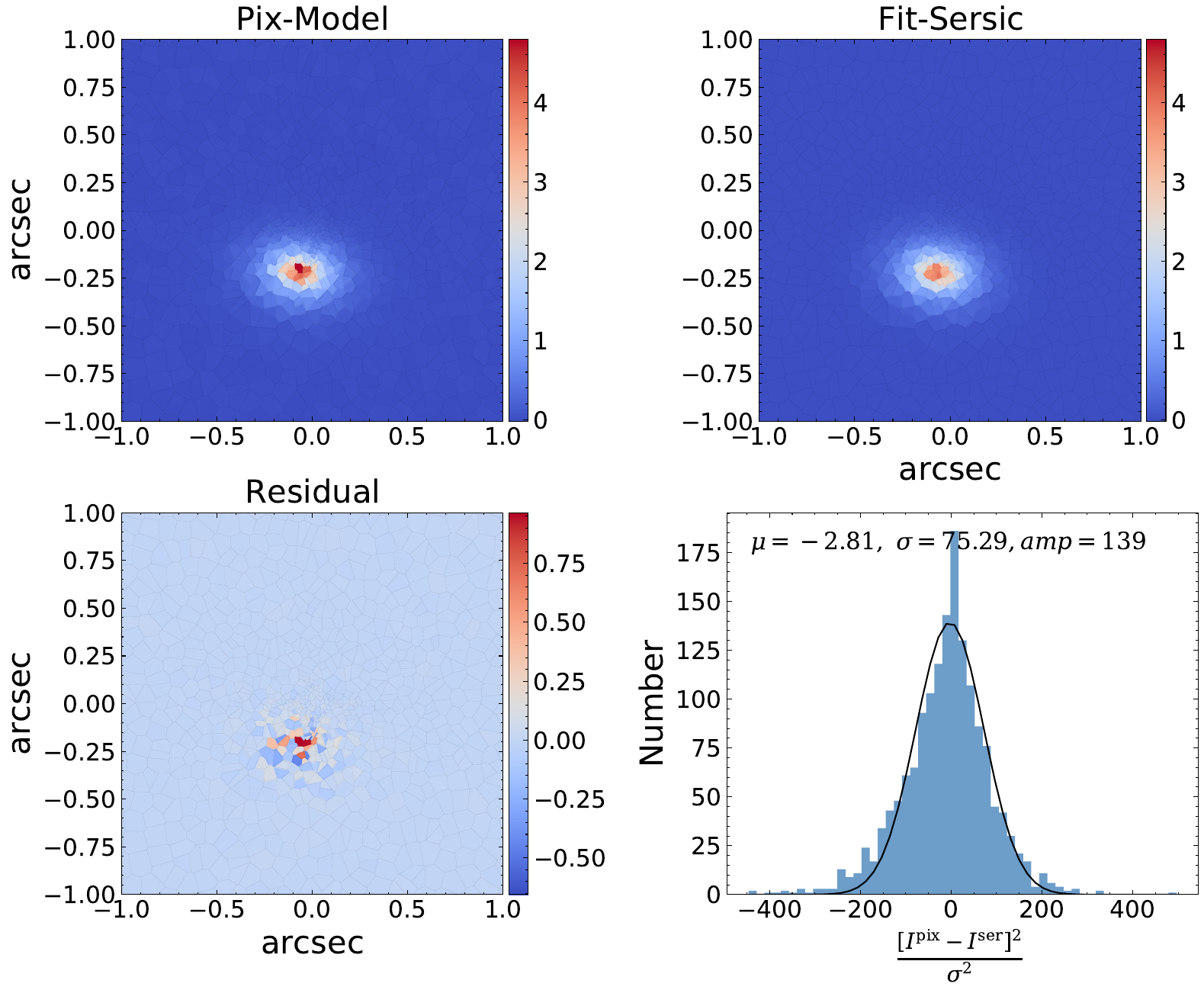}
    \caption{The source reconstruction of the MaNGA lens ``1344\_8263-9102''. Top-left: the pixelized source reconstruction corresponding to the best-fit lens model. Top-right: the best-fit Sersic model to the pixelized source reconstruction, shown using the same Voronoi pixels as used in the source reconstruction. Bottom-left: the residual map of the best Sersic fit, i.e, top-left image minus top-right image. \reply{Bottom-right: the blue histogram shows the number distribution of the normalized residual ($\frac{[I^{\rm pix}-I^{\rm ser}]^2}{\sigma^2}$, see equation~\eqref{eq:source_fitting}) of the best Sersic fit. A Gaussian fit to this distribution is shown with the black line, the corresponding best-fit mean ($\mu$), standard deviation ($\sigma$), amplitude value ($amp$) are shown in the text label. This distribution has a tail with large normalized residual values ($\sim100$), because the source uncertainties are typically underestimated in the outskirts of the source.}
    }
    \label{fig:5_src_typical_case}
\end{figure*}

The source morphology is also typically well recovered. To compare the pixelized model inferred by {\tt PyAutoLens} to the input Sersic model, we fit the model source-plane image with another Sersic, by minimizing the following $\chi^2$ function
\begin{equation}
\chi^2 = \sum_{i} \frac{[I^{\rm pix}_{i}-I^{\rm ser}_{i}]^2}{\sigma_{i}^2},
\label{eq:source_fitting}
\end{equation}
where $I^{\rm pix}_{i}$, and $I^{\rm ser}_{i}$ are the pixelized and Sersic model intensities at pixel $i$ in the source plane, respectively, and $\sigma_{i}$ is the corresponding source intensity uncertainties given by the lens model \citep{Warren03,Suyu06}. The fit for a typical lens systems is shown in Figure~\ref{fig:5_src_typical_case},  and the best-fit Sersic parameters are compared to the truth in Table \ref{tab:1_source_tab}.
For this lens system, we infer the axis-ratio, position angle, and Sersic index of the source without any obvious bias. The relative systematic errors on the axis-ratio and Sersic index are 0.2\% and 1.5\% respectively, while the systematic error on the position angle is only -0.18 degrees. The half-light radius is marginally underestimated (3.6\%). This underestimation of the source size is predicted by the SPT and is consistent with the convergence error map shown in Figure~\ref{fig:4_typical_case_model_result}, specifically in the regions where the strongly lensed images are located. The convergence field of the model ($\kappa_\mathrm{model}$) is larger than the true value ($\kappa_\mathrm{data}$), such that the lens model's $\kappa_\mathrm{model}$ can be qualitatively understood as a re-scaled $\kappa_\mathrm{data}$ with the addition of a \reply{positive} mass sheet. This corresponds to $\rm \lambda<1$ in equation~\eqref{eq:1_MST}, so the source size given by the model shrinks.

Two possibilities might explain why systematic errors are larger than statistical error.
First, the model is over-constrained because we assume an EPL mass model with insufficient complexity to represent our mock ``MaNGA lenses''. Any restricted mass model drives a precise but inaccurate source reconstruction \citep{Kochanek20}. Second, a recent study has shown that the advanced pixelized source modelling provided by {\tt Pyautolens} leads to stronger discretization noise comparing with other ``traditional'' methods \citep[e.g.][]{Suyu06,lmass_Vegetti09} without ``random pixelization'' features \citep[see][]{Nightingale15}, which may also weakly bias the source reconstruction. A ``stochastic likelihood cap'' can mitigate this kind of bias (Etherington et al.\ in prep.).

\subsubsection{Outliers}
\label{sec:source_outliers}
Except for the typical system shown in Section~\ref{sec:source_typical_case}, we notice that there are some ``outliers'' in our modeling results. Although the model still roughly recovers the source's shape (axis-ratio, position angle), the source's size is significantly misestimated. To make sure these ``outliers'' are not caused by bad lens modeling, we visually inspect the normalized residual map\footnote{\url{https://github.com/caoxiaoyue/Sim_MaNGA_lens/tree/main/best_fit_result_subplot}} (bottom-left panel in Figure~\ref{fig:4_typical_case_model_result}) and the source reconstruction image\footnote{\url{https://github.com/caoxiaoyue/Sim_MaNGA_lens/tree/main/src_fit_image}} (top-right panel in Figure~\ref{fig:5_src_typical_case}) of each lens system. We find the normalized residuals are typically within the $\sim 3 \sigma$ limit and that every source reconstruction appears physical, i.e., it does not show the unphysical structure that is demonstrated in \citet{unphy_src_Maresca21}. Therefore, our lens modeling results are fairly good.

\begin{figure*}
	\includegraphics[width=\textwidth]{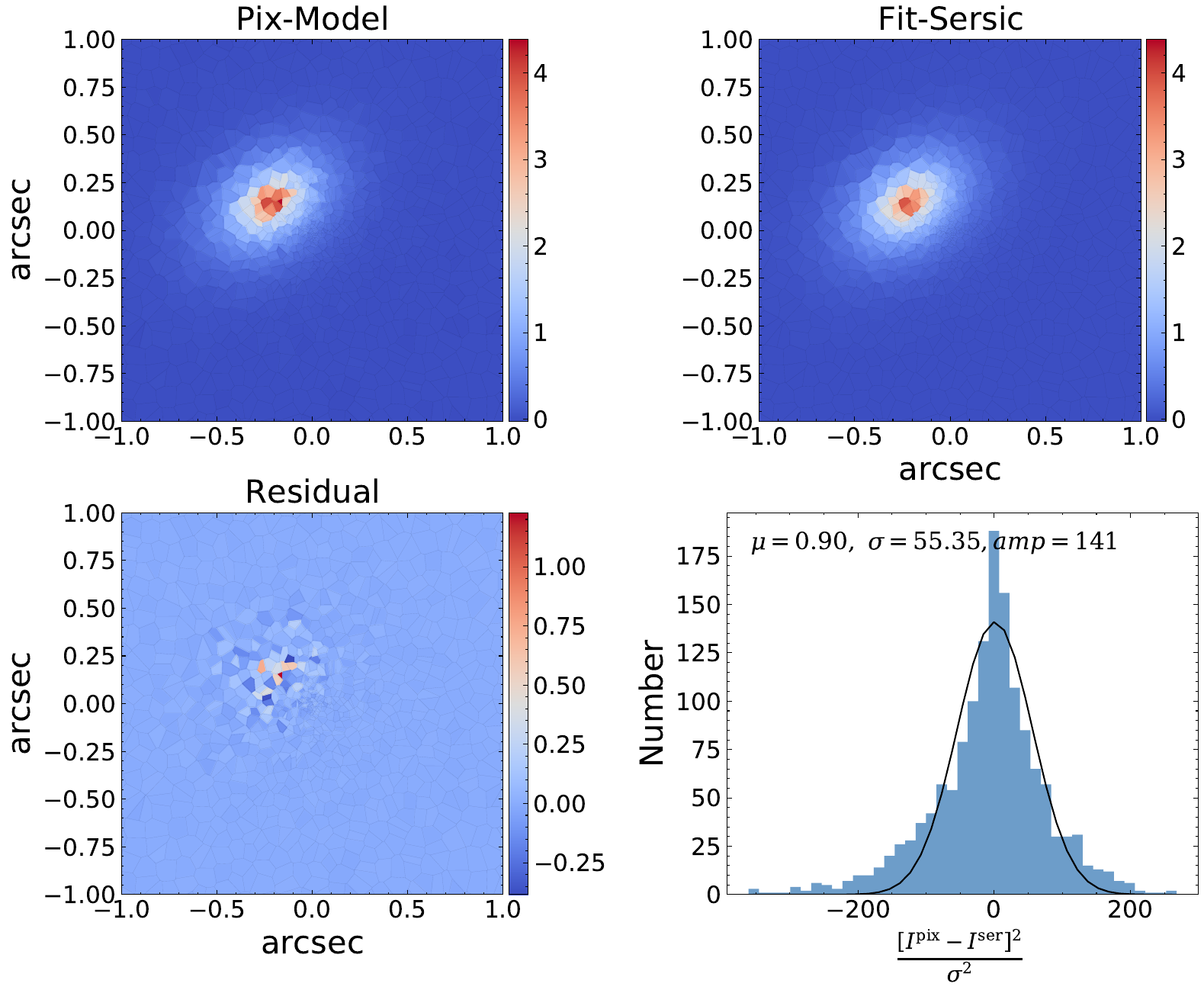}
    \caption{Similar to Figure~\ref{fig:5_src_typical_case}, the source reconstruction of the MaNGA lens ``3485\_8153-3701''.}
    \label{fig:6_src_outliers}
\end{figure*}

The cause of outliers is essentially due to the mismatch between the ideal EPL model and the ``true'' mass distributions of our mock lenses, which results in a biased source reconstruction via the SPT. Figure~\ref{fig:6_src_outliers} shows a representative outlier (``3485\_8153-3701''), whose modeling bias can be quantitatively understood via the MST (a special case of the SPT). Recall the mass profile of the lens system ``3485\_8153-3701'' we presented in Figure~\ref{fig:7_outliers_avg_kappa_prof}, the green dashed line represents the mass-sheet transformed solution of the deflector's true mass profile. The corresponding $\lambda$ factor is given by the ratio of the source size inferred by the lens model to the input true value. The coincidence of the green dashed, orange dashed and red solid line in the shaded grey region indicates that the degeneracy in this system can be well understood via the MST. More specifically, the $\lambda$ factor corresponding to the orange dashed line is 1.42, which is consistent with the model's source size being overestimated by a factor of 1.49 (recall Equations~\ref{eq:1_MST} and \ref{eq:2_MST}).

%----------------------figure and table begin

\begin{table*}
	\centering
	\caption{
	A Sersic model is used to fit the pixelized source-reconstruction of our mock ``MaNGA lenses''. The best-fit parameters are compared to the input source parameters that generated the data, allowing us to quantify how much the mass model mismatch bias the source morphology.
	Column 0 gives the identifier of our ``MaNGA lenses''. Columns 1 and 2 list the source parameter names (units) and corresponding best Sersic-fit values ($\rm median  \pm 1\sigma$ errors). The input true values of source parameters are given in Column 3. 
}
	\label{tab:1_source_tab}
\begin{tabular}{|c|c|c|c|} %@{\vline}
\hline
Lens identifier & Parameters\;(unit) & $\rm Median \pm Error$ & Ground truth  \\
{\scriptsize (1)} & {\scriptsize (2)} & {\scriptsize (3)} & {\scriptsize (4)} \\
\hline
\multirow{4}{*}{1344\_8263-9102} &$r_\mathrm{e}$\;(\arcsec) & $0.144530 \pm 0.000075$ & 0.15 \\ 
%\cline{2-4} 
                  &$n_\mathrm{s}$ & $0.985375 \pm 0.000714$ & 1.0 \\ %\cline{2-4} 
                  &$q_{\rm s}$ & $0.701550 \pm 0.000078$ & 0.7  \\ 
                  %\cline{2-4} 
                  &$\phi_{\rm s}$\;($^{\circ}$) & $172.238155 \pm 0.008920$ & 172.420115 \\ 
\hline
\multirow{4}{*}{3485\_8153-3701} &$r_\mathrm{e}$\;(\arcsec)  & $0.223250 \pm 0.000054$  & 0.15  \\ 
%\cline{2-4} 
                  &$n_\mathrm{s}$  & $0.960752 \pm 0.000370$  & 1.0  \\
                  %\cline{2-4} 
                  &$q_{\rm s}$  & $0.693070 \pm 0.000059$  & 0.7  \\ 
                  %\cline{2-4} 
                  &$\phi_{\rm s}$\;($^{\circ}$) & $34.556441 \pm 0.006671$ & 35.742580   \\ 
\hline
\end{tabular}
\end{table*}

%----------------------figure and table end

\subsubsection{Population statistics}
\label{sec:source_stat}
\begin{figure}
	\includegraphics[width=\columnwidth]{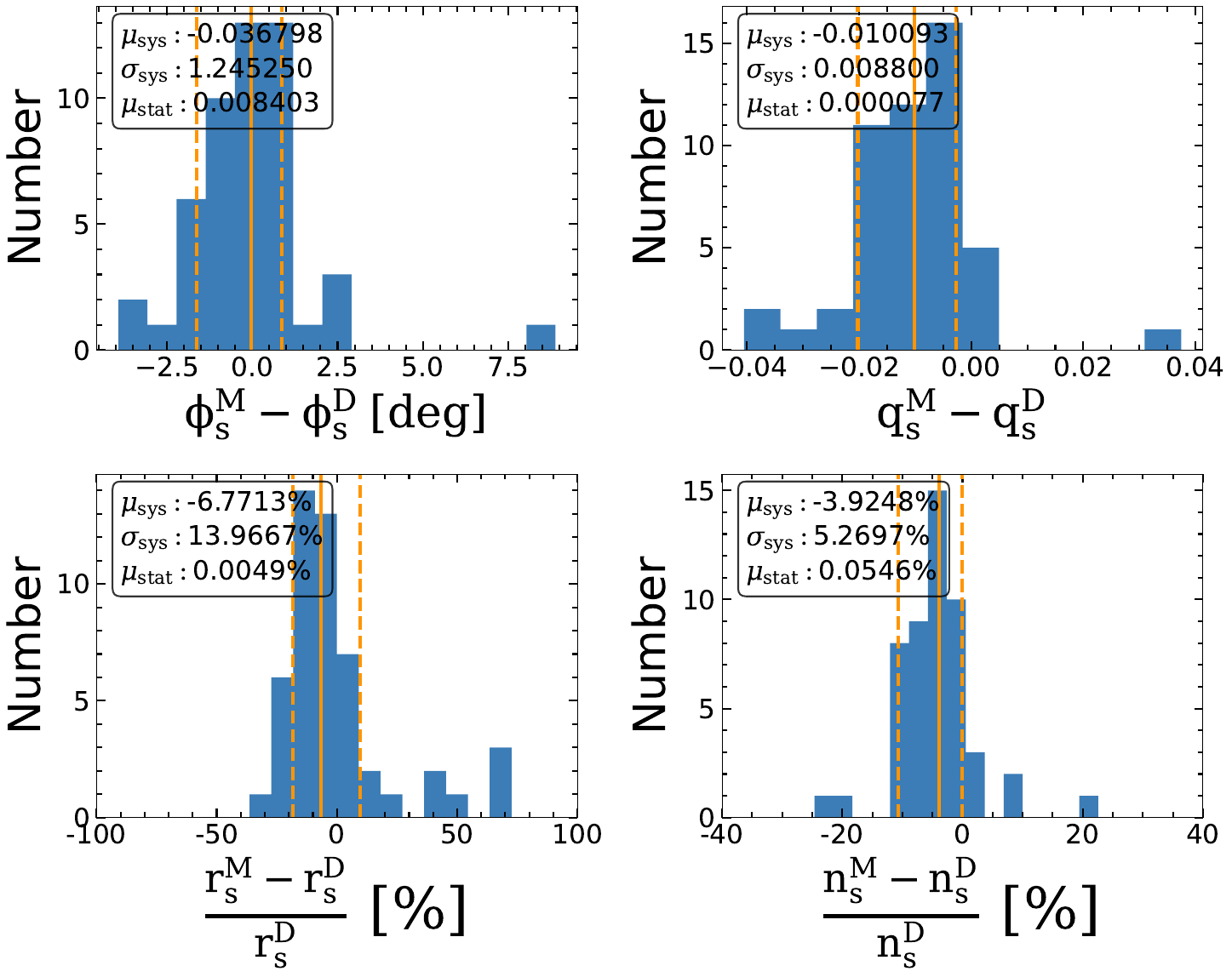}
    \caption{
    The top panels show the distribution of absolute systematic errors for sources' position angle ($\rm \phi_{\rm s}$) and axis-ratio ($\rm q_{\rm s}$), with the median ($\rm \mu_{\rm sys}$) and standard deviation (or dispersion, $\rm \sigma_{\rm sys}$) of the systematic error, and the median value of the statistical error ($\rm \mu_{\rm stat}$ ,purely predicted by the non-linear search sampler) overlay on the legend. The bottom panels show the distribution of relative systematic errors for sources' effective radius ($\rm r_{\rm s}$) and Sersic index $\rm n_{\rm s}$. Three orange lines mark the [16\%, 50\%, 84\%] percentiles of the histograms.
    }
    \label{fig:src_all_in_one}
\end{figure}

For all 50 ``MaNGA lenses'' simulated in this work, we carry out the analysis described in Section~\ref{sec:source_typical_case}. The global morphology properties of our pixelized source reconstructions are statistically summarized in Figure~\ref{fig:src_all_in_one}. We find that the lens models accurately recover the sources' position angles ($\phi_s$, absolute systematic errors with the median value of $-0.0368$ degrees, and standard deviation of $1.2452$ degrees) and axis-ratio ($q_s$, absolute systematic errors with the median value of $-0.0101$ and standard deviation $0.0088$); However, the size of the source ($r_s$) and its Sersic index ($n_s$) are underestimated by 6.77\% and 3.92\% on average. Combining the result of both $r_s$ and $n_s$, sources reconstructed by the lens model are slightly more compact than their true appearance.

\begin{figure}
 	\includegraphics[width=\columnwidth]{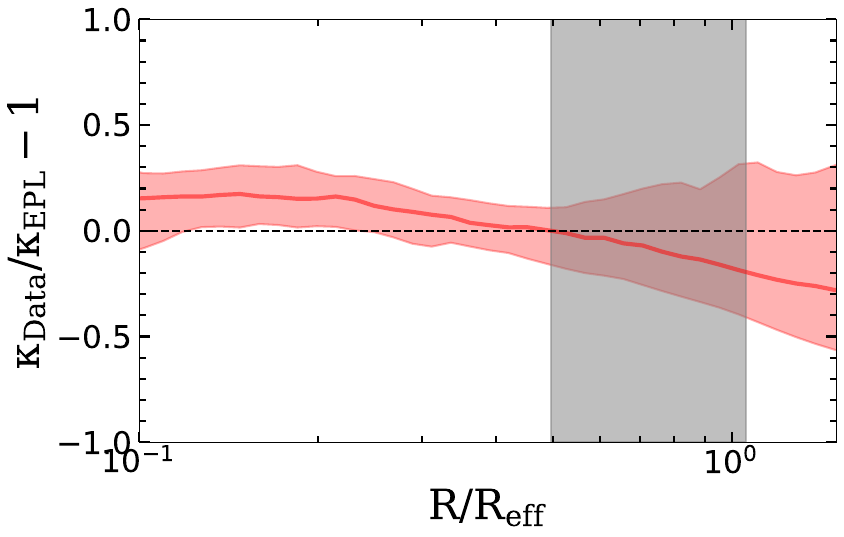}
    \caption{The residual projected density profile is plotted as a function of radius (normalised by the effective radius of the lens's light). This shows the difference between the best-fit model's surface density and the ground truth values. This was calculated for each of our 50 mock lenses, with the red solid line showing the median deviation between the best-fit models and the respective truths, while the red shaded region illustrates the 16th-84th percentiles of the residuals at each radius.The grey shaded region represents the 16th--84th percentiles of the distribution of $\theta_\mathrm{E}/R_\mathrm{eff}$ in our lens sample. The black dashed line represents the curve when the lens model perfectly recovers the surface density profile that generated the data.}
    \label{fig:8_kappa_1d}
\end{figure}

\begin{figure}
	\includegraphics[width=\columnwidth]{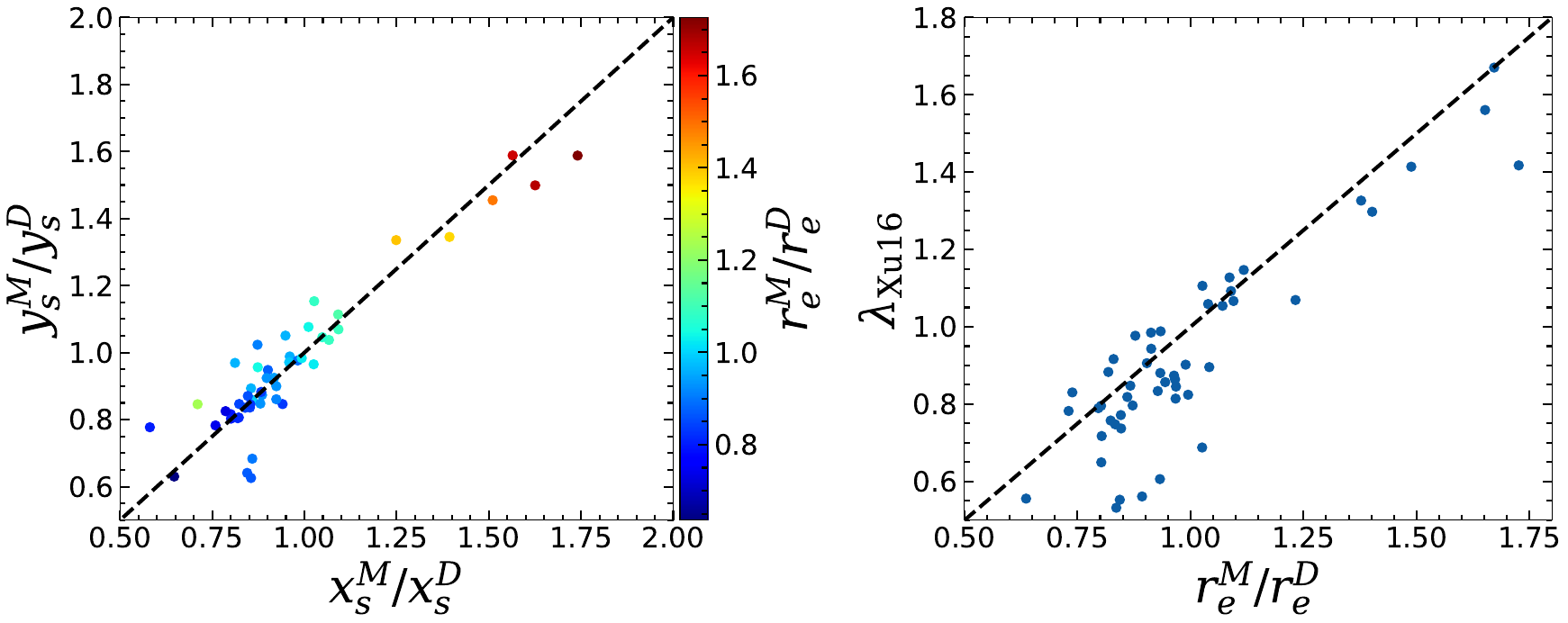}
    \caption{\reply{
Left panel: $x_s^M, y_s^M, r_e^M$ represents the position of the source, and its half-light radius, inferred by the lens model. $x_s^D, y_s^D, r_e^D$ corresponds to the input true value of the mock data. Each dot corresponds to a lens system, which is usually found around the black dotted line that represents a 1:1 relationship. This means that for most lens systems, the position of the source is rescaled according to the relation $(x_s^M, y_s^M)=(\lambda x_s^D, \lambda y_s^D)$; The size of the source (coded by the color bar) also roughly matches the relationship $r_e^M = \lambda r_e^D$. Right panel: for most lens systems, the $\lambda$ factor given by the change of the source size ($\frac{r_e^M }{r_e^D}$) is roughly consistent with the $\lambda$ value derived by equation~\eqref{eq:mst_lambda} ($\lambda_{\rm Xu16}$). Combining the left and right panels, we conclude that the lensing degeneracy in most lens systems manifests itself in an MST fashion.}
    }
    \label{fig:check_mst}
\end{figure}

Our algorithm tends to reconstruct an overly compact source galaxy on average because the input mass distributions are cuspier than can be described by an EPL model (Figure~\ref{fig:8_kappa_1d}).This mismatch leads to $\sim 7\%$ underestimation of the convergence near the Einstein ring, and $\sim$ 7\% underestimation of the source size via the SPT.
\footnote{When lensing degeneracies are manifested as the MST, The underestimation of the convergence near the Einstein radius is indirectly related to the underestimation of the source size through the $\rm \bar{\kappa}(<\theta)$ (see Equation~\eqref{eq:4_alpha_kappa_mean}). Therefore, the consistent $\sim$7\% number here could be just a coincidence.}
\reply{We find the degeneracy in most of lens systems can be approximated as an MST (see Figure~\ref{fig:check_mst})}. Of our 50 mock lens systems, the sizes of 39 source galaxies are modeled within $20\%$ of the true value, the remaining 11 systems we designate as ``outliers''. Among the ``outliers'', the modeling bias in 9 systems can be understood in terms of the MST (see the discussion in Section~\ref{sec:source_outliers}); The remaining 2 outliers reflect the more general SPT, where the difference between the model and the true value of the two-dimensional convergence map in the region of the lensed arc cannot be compensated by a uniform mass sheet, in either the radial or angular direction. We further show those 2 ``Non-MST outliers'' in Appendix~\ref{sec:appdix_nmst}.

%------------------------time-delay
\subsection{Time delay and $H_0$ inference}
\label{sec:time_delay}
\begin{figure}
	\includegraphics[width=\columnwidth]{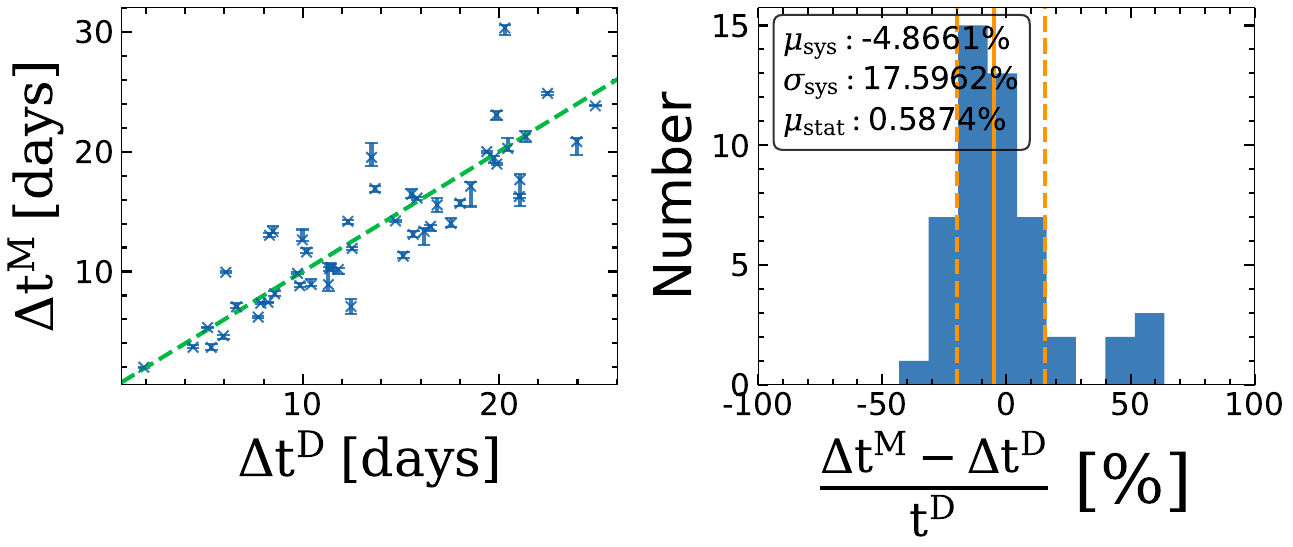}
    \caption{
    Reconstructed measurements of the time-delay between pairs of images of the same source, $\Delta t^\mathrm{M}$, across our mock lens sample, compared to the true value, $\Delta t^\mathrm{D}$.
    These are underestimated by $\sim5\%$ on average.
    }
    \label{fig:13_td}
\end{figure}
For every pair of multiply lensed images (marked by green circles in Figure~\ref{fig:1_mock_lens_image}), we calculate the relative time delay defined in equation~\ref{eq:rel_td}). By comparing the model value of each pairwise relative time delay $\rm \Delta t^M$ with its true value $\rm \Delta t^D$ (Figure~\ref{fig:13_td}), we investigate the systematic error caused by our assumption of the EPL + shear model. We find relative time delays to be underestimated, with median relative systematic error 5\% and standard deviation 18\%. This underestimation is qualitatively consistent with the underestimation of $\kappa$ near the Einstein radius for most lenses in our sample (see the average trend in Figure~\ref{fig:8_kappa_1d}), which rescales the relative time delay according to \eqref{eq:3_MST}. Furthermore, the 6 lens systems whose time delay measurements are discrepant by more than 4\,days\footnote{Considering the typical time-domain measurement uncertainty for current instruments is $\rm \sim 2\,days$, 4\,days roughly corresponds to $2 \sigma$ error.}, also suffer from significantly overestimated or underestimated $\kappa$ at the image locations.
Thus, although the EPL model can approximately describe the density profile of the lens in a statistical sense, the density profile of an individual lens system may still deviate significantly from the EPL model, resulting in large discrepancies between the predicted and true relative time delays.

\begin{figure}
	\includegraphics[width=\columnwidth]{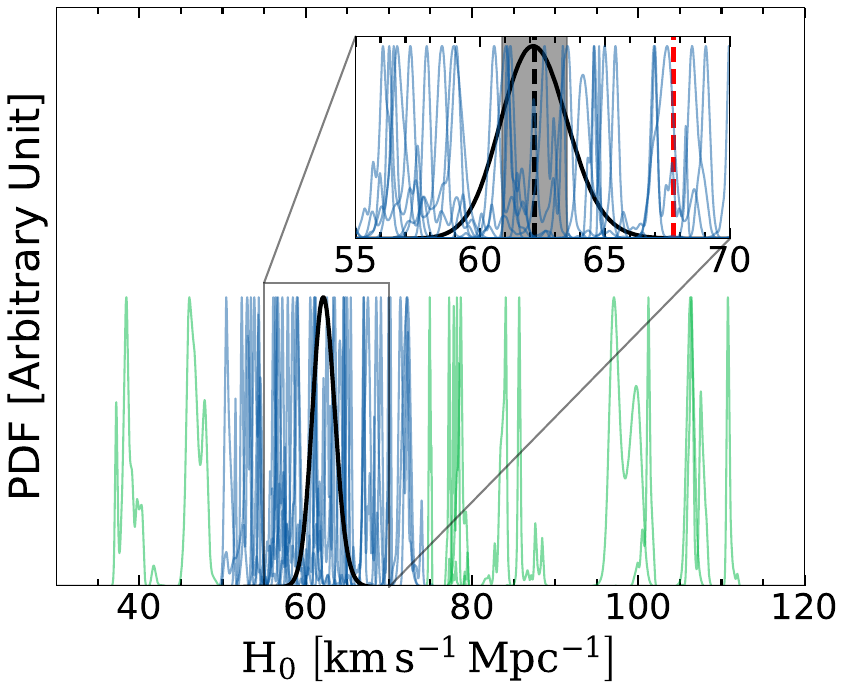}
    \caption{Probability density function (PDF) of $H_\mathrm{0}$ predicted by each individual lens, shown with blue and green histograms. The width of each blue and green histogram is purely governed by the statistical errors of the lens mass model. Lenses corresponding to green histograms are excluded from the joint $H_\mathrm{0}$ inference via the $2 \sigma$ clipping, to avoid the bias introduced by these extreme systems. The black histogram is the PDF of $H_\mathrm{0}$ constrained by a joint analysis of 37 lenses (indicated by blue histograms). We re-normalize each PDFs to the same height for better visualization. The inset plot shows the zoomed-view around the black histogram. The red vertical dashed line marks the input true value of the $H_\mathrm{0}$. The black vertical dashed line and the shaded grey mark the median value and the 68\% confidence interval of the black histogram (i,e, $\rm 62.3 \pm 1.3 \; km\,s^{-1}\,Mpc^{-1}$); this prediction of $H_0$ is $\sim 4\sigma$ lower than the input true value ($\rm67.7 \; km\,s^{-1}\,Mpc^{-1}$).
    }
    \label{fig:14_H0_ensemble}
\end{figure}

In time-delay cosmography, a biased estimation in the relative time delay between pairwise lensed images can be further interpreted as a biased estimation of Hubble constant ($H_0^{\rm M}$), via the following relation (see the derivation in appendix~\ref{sec:h0_td})
%used to infer the Hubble constant
 \begin{equation}
H_0^\mathrm{M} = \left(\frac{\Delta t^\mathrm{M}}{\Delta t^\mathrm{D}}\right) H_0^\mathrm{D}~,
\end{equation}
where we assume a fiducial value $H_0^\mathrm{D}=67.7\,$km\,s$^{-1}$\,Mpc$^{-1}$. Figure~\ref{fig:14_H0_ensemble} shows the probability density function (PDF) of $H_0$ obtained using the inferred pairwise relative time delays of every lens system. The width of each PDF is purely governed by the statistical errors of the lens mass model. They are narrow because the over-simplified EPL mass model can drive a very precise but inaccurate lens mass model \citep{Kochanek20}. We find $H_0^\mathrm{M}$ measurements randomly distributed between $\rm 40\,km\,s^{-1}\,Mpc^{-1}$ and $\rm 110\,km\,s^{-1}\,Mpc^{-1}$. This result is significantly less homogeneous than state-of-art time-delay cosmography projects. For example, the H0LICOW project measures $H_0$ using 6 gravitationally lensed quasars with measured time delays, and 5 of them are analyzed blindly \citep{wong20}. In their results, the $H_0$ values predicted by each lens cluster around $\rm 73.3\,km\,s^{-1}\,Mpc^{-1}$, and yield statistically consistent results (see Figure~2 in \citet{wong20}), indicating that they do not underestimate statistical errors and successfully control the systematics. Systematic errors are larger in our results, because our pure image-based lens model is subject to degeneracies of a ``pseudo internal mass-sheet'' (see Section~\ref{sec:Lensing_Degeneracy}), induced by the mismatch between the ideal EPL model and the more complex distribution of mass in our mock ``MaNGA lenses''. To reduce this systematic and obtain an unbiased $H_0$ measurement, velocity dispersion is the crucial information \citep{stellar_dynamics_H0_analysis}.

We further explore using multiple lens systems to jointly constrain $H_0$. We write the joint likelihood as
\begin{equation}
\label{eq:like_dt}
 P(\boldsymbol{\Delta \rm t} | H_0) = \\ \prod_{i} \left( \frac{1}{\sqrt{2\pi}\sigma_{{\Delta \rm t},i}} \exp \left[ \frac{({\Delta \rm t}_i^{\rm D}-{\Delta \rm t}_i^{\rm M}(H_\mathrm{0}))^2} {2\sigma_{{\Delta \rm t},i}^2} \right] \right)~,
\end{equation}
where the subscript $i$ runs over all lens systems. We assume uncertainty on time delay measurements $\sigma_{{\Delta \rm t},i}=2$\,days, which is typical for recent time-domain observations, and adopt a uniform prior for $H_0$ from 0 to $\rm 150\, km\,s^{-1}\,Mpc^{-1}$. Results of a joint inference on $H_0$ are shown in Table-\ref{tab:H0_tab} and Figure~\ref{fig:14_H0_ensemble}. Including all lens systems, or with $3\sigma$ clipping, we obtain measurements of $H_0$ whose systematic biases are below the $\sim 1\sigma$ level. However, this result is driven by outliers (green curves) and is inconsistent with the mass profile trend (Figure~\ref{fig:8_kappa_1d}) which suggested that $H_0$ should be underestimated by $\sim 10$\%. Aggressive $2\sigma$ clipping efficiently removes unreliable measurements of $H_0$. A joint analysis of the remaining 37 systems yields a posterior distribution of $H_0$ (black curve) centered at $\rm 62.25 \pm 1.46 km\,s^{-1}\,Mpc^{-1}$. This is underestimated by 9\% and $\sim4\sigma$ lower than the input truth, but consistent with the mass profile trend (Figure~\ref{fig:8_kappa_1d}).

We interpret the bias in $H_0$ measurements to be dominated by the mass mismatch between the EPL model and complex real lenses, coupled with the MST. There are two possible routes to reduce this kind of bias. First, better-matched lens mass parametrizations could avoid the pseudo-internal mass sheet entirely (see Section~\ref{sec:Lensing_Degeneracy}). Second, measurements of stellar dynamics, which provide 2D enclosed mass at a radius that is different from the Einstein radius, would avoid the bias by breaking the degeneracy (specifying the value of the pseudo-internal mass sheet). Recently, \cite{Birrer20} applied an internal mass sheet in combination with an EPL model to produce a more flexible lens mass model, where the strength of the internal mass sheet is determined exclusively by stellar dynamics on the population level. Testing this methodology on mock lenses generated from the cosmological hydrodynamics simulation \citep{Ding21} shows that an unbiased H0 measurement can be obtained.

\begin{table}
	\centering
	\caption{$H_0$ constraints given by the joint analysis. Column 2 shows the number of lens systems that are used in joint analyses. The lens systems are selected with clipping schemes shown in Column 1 (no-clipping, $\rm 3\sigma$ clipping, $\rm 2\sigma$ clipping). Prediction values of $H_0$ and corresponding systematic errors relative to the ground truth ($\rm67.7 \; km\,s^{-1}\,Mpc^{-1}$) are listed in Columns 3 and 4, respectively.}
	\label{tab:H0_tab}
\begin{tabular}{|c|c|c|c|}
\hline
Clip Type  & $\rm N_{\rm lens}$ & $H_0$\;[$\rm km\,s^{-1}\,Mpc^{-1}$]   & Bias  \\
{\scriptsize (1)} & {\scriptsize (2)} & {\scriptsize (3)} & {\scriptsize (4)} \\
\hline
No clip    & 50  & $68.96\pm1.26$ & $0.87\sigma$  \\
$\rm 3 \sigma$\;clip & 49  & $68.77\pm1.38$ & $0.75\sigma$  \\
$2 \sigma$\;clip & 37  & $62.25\pm1.46$ & $-3.78\sigma$ \\
\hline
\end{tabular}
\end{table}

\section{Conclusion and summary}
\label{sec:conclution}
This work provides a new alternative for testing the systematic error of lens modeling. We use the results of dynamic mass reconstruction of MaNGA ETGs to simulate mock lenses that are similar to the SLACS project. We expect that our test suite reflects many of the complexities of real observational lensing data, such as radial density profiles that deviate from a power-law and angular structures that do not possess elliptical symmetry. Comparing with previous similar works which are typically based on cosmological hydrodynamics simulations, our dataset avoids the limitations of current simulations (such as insufficient particle resolution and imperfect subgrid physics). We find that in a statistical sense: 
\begin{enumerate} 
\item{The Einstein radius can be well recovered with 0.1\% accuracy for our SLACS-like mock lenses.} 
\item{The non-elliptical symmetry of a lens can produce a certain level of internal shear measurement, with an average strength of $\sim$0.015.}
\item{lensing fails to measure the deflector's local density slope in the region of the lensed emission, due to the SPT induced by the mass distribution mismatch between the ideal EPL model and real lenses. As a comparison, we find the deflector's global density slope is better measured; The relative systematic error has a median value of -2.5\%, with a 3.4\% intrinsic scattering. This result demonstrates by fully exploring the vast information in the extended arc, the pure image-based lens modeling can measure the deflector's global density slope with enough accuracy to guide the galaxy formation and evolution theory.
}
\item{The global morphology of the source can be reconstructed reasonably well (such as axis ratio, position angle, half-light radius, and Sersic index), although the sources predicted by lens modelling are typically slightly more compact than the truth.}
\item{Using the pure image-based lens modeling with extended arcs, the $H_0$ value inferred from a time-delay measurement in a single lens has a large random systemic error, with inferred values typically ranging from 40 to 110 $\rm km\,s^{-1}\,Mpc^{-1}$. Combining the model results of 37 lens systems, the random systematic errors that exists in the individual lenses cancel at some level, a joint analysis of 37 lenses under-predicts the $H_0$ value by $\sim9\%$ and with a $\sim 4\sigma$ bias. This underestimation is consistent with the trend that the lens model tends to slightly under-predict the convergence value in the vicinity of the Einstein radius. The next-generation time-delay cosmography typically hopes to suppress the systematic error to below $\sim 1\%$; in order to achieve this goal, a better form of mass parametrization and additional information (such as the stellar dynamics) are required to further break the degeneracy we discussed in Section~\ref{sec:mass_slope}.}
\end{enumerate}
\reply{For a large number of lenses, lens modeling reliably gives the population distribution of source and lens parameters. However, we emphasize this result is for a particular set of MaNGA galaxies, and the specific value of the detected biases may change for another set of strong lenses.}

For an individual lens, its mass profile may deviate from the form of the EPL model significantly, which leads to large systematic errors in the lens modeling results and becomes the ``outliers''. However, due to the source position transformation, we may not be able to judge whether a lens is an ``outlier'' by checking the goodness of fit of modeling results (such as the $\chi^2$ map). To some extent, the existence of ``outliers'' reduces the reliability of the model prediction given by an individual lens.

By breaking the lens modeling procedure into a series of simpler model fits with {\tt PyAutoLens}, our model pipelines are fully automated. This allows us to apply our current methods to future wide-field observations, such as CSST \citep{csst1}, Euclid \citep{euclid}, and LSST \citep{lsst}, where more than one hundred thousand lenses are expected to be discovered \citep{Collett15}.
%-----------------------------
Although our works are based on relatively ideal mock samples, conclusions we got may still guide the application of future wide-field lensing observations.

Recently, several studies have shown that the angular structure of lenses also plays an important role in $H_0$ estimation. \citet{Gomer21} found that the surface mass density of real-life lenses has the component that deviating from the elliptical symmetry, which can be approximately described by an additional dipole term. If the EPL + shear mass model is used to do the lens modeling, the systematic error of $H_0$ prediction can reach $\sim 10\%$. \citet{Kochanek21} points out that if the mass model used in lens modeling does not have sufficient degrees of freedom in the angular direction, the constraint from the Einstein ring on the lens's angular structure drives the selection of radial density profile, resulting in a very precise but inaccurate $H_0$ prediction. The dynamic reconstruction of ETGs given by the JAM model is more realistic in angular direction (i.e, is not perfectly elliptical-symmetric), which makes our mock sample an ideal playground to test the robustness of the current lens modeling pipeline, especially for the purpose of $H_0$ inference. We hope this work can benefit the community, all the mock data and its generation scripts used in this work are publicly available from the \url{https://github.com/caoxiaoyue/Sim_MaNGA_lens}.

\section*{Software Citations}

This work uses the following software packages:

\begin{itemize}

\item
\href{https://github.com/astropy/astropy}{{Astropy}}
\citep{astropy1, astropy2}

\item
\href{https://bitbucket.org/bdiemer/colossus/src/master/}{{Colossus}}
\citep{colossus}

\item
\href{https://github.com/dfm/corner.py}{{Corner.py}}
\citep{corner}

\item
\href{https://github.com/joshspeagle/dynesty}{{Dynesty}}
\citep{dynesty}

\item
\href{https://github.com/dfm/emcee}{{Emcee}}
\citep{emcee}

\item
\href{https://github.com/sibirrer/lenstronomy}{{Lenstronomy}}
\citep{lenstro01,lenstro02,lenstro03}

\item
\href{https://github.com/matplotlib/matplotlib}{{Matplotlib}}
\citep{matplotlib}

\item
\href{numba` https://github.com/numba/numba}{{Numba}}
\citep{numba}

\item
\href{https://github.com/numpy/numpy}{{NumPy}}
\citep{numpy}

\item
\href{https://github.com/rhayes777/PyAutoFit}{{PyAutoFit}}
\citep{pyautofit}

\item
\href{https://github.com/Jammy2211/PyAutoLens}{{PyAutoLens}}
\citep{Nightingale15, Nightingale18, pyautolens}

\item
\href{https://github.com/equinor/pylops}{{PyLops}}
\citep{pylops}

\item
\href{https://github.com/JohannesBuchner/PyMultiNest}{{PyMultiNest}}
\citep{multinest, pymultinest}

\item
\href{https://github.com/jyhmiinlin/pynufft}{{PyNUFFT}}
\citep{pynufft}

\item
\href{https://github.com/AshKelly/pyquad}{{Pyquad}}
\citep{pyquad}

\item
\href{https://github.com/ljvmiranda921/pyswarms}{{PySwarms}}
\citep{pyswarms}

\item
\href{https://www.python.org/}{{Python}}
\citep{python}

\item
\href{https://github.com/scikit-image/scikit-image}{{Scikit-image}}
\citep{scikit-image}

\item
\href{https://github.com/scikit-learn/scikit-learn}{{Scikit-learn}}
\citep{scikit-learn}

\item
\href{https://github.com/scipy/scipy}{{Scipy}}
\citep{scipy}

\end{itemize}

\normalem
\begin{acknowledgements}
\reply{We thank the referee for the thoughtful comments
and suggestions that improved the organization of the manuscript.} RL acknowledges the support of the National Nature Science Foundation of China (Nos 11773032, 12022306), the science research grants from the China Manned Space Project (No CMS-CSST-2021-B01). JWN and RJM acknowledge funding from the UK Space Agency through award ST/W002612/1, and from STFC through award ST/T002565/1.
CSF and AA acknowledge support by the European Research Council (ERC) through Advanced Investigator grant to CSF, DMIDAS (GA 786910).
AR is supported by the European Research Council Horizon2020 grant `EWC' (award AMD-776247-6). 
NCA is supported by an STFC/UKRI Ernest Rutherford Fellowship, Project Reference: ST/S004998/1.
\end{acknowledgements}
  
\bibliographystyle{raa}
\bibliography{spt}

\begin{thebibliography}{155}
\providecommand\natexlab[1]{#1}
\providecommand\JournalTitle[1]{#1}

\bibitem[{Amiaux} {et~al.}(2012)]{euclid}
{Amiaux}, J., {Scaramella}, R., {Mellier}, Y., {et~al.} 2012, in Society of
  Photo-Optical Instrumentation Engineers (SPIE) Conference Series, Vol. 8442,
  Space Telescopes and Instrumentation 2012: Optical, Infrared, and Millimeter
  Wave, ed. M.~C. {Clampin}, G.~G. {Fazio}, H.~A. {MacEwen}, \& J.~{Oschmann},
  Jacobus~M., 84420Z

\bibitem[{Astropy Collaboration} {et~al.}(2013)]{astropy1}
{Astropy Collaboration}, {Robitaille}, T.~P., {Tollerud}, E.~J., {et~al.} 2013,
  \aap, 558, A33

\bibitem[Auger {et~al.}(2009)]{Auger09_stellar}
Auger, M.~W., Treu, T., Bolton, A.~S., {et~al.} 2009, The Astrophysical
  Journal, 705, 1099–1115

\bibitem[{Auger} {et~al.}(2010)]{lmass_Auger10_lens}
{Auger}, M.~W., {Treu}, T., {Bolton}, A.~S., {et~al.} 2010, \apj, 724, 511

\bibitem[{Birrer} \& {Amara}(2018)]{lenstro02}
{Birrer}, S., \& {Amara}, A. 2018, {Lenstronomy: Multi-purpose gravitational
  lens modeling software package}

\bibitem[{Birrer} {et~al.}(2015)]{lenstro01}
{Birrer}, S., {Amara}, A., \& {Refregier}, A. 2015, \apj, 813, 102

\bibitem[{Birrer} {et~al.}(2018)]{shear_birrer18}
{Birrer}, S., {Refregier}, A., \& {Amara}, A. 2018, \apjl, 852, L14

\bibitem[{Birrer} {et~al.}(2017)]{shear_birrer17}
{Birrer}, S., {Welschen}, C., {Amara}, A., \& {Refregier}, A. 2017, \jcap,
  2017, 049

\bibitem[{Birrer} {et~al.}(2020)]{Birrer20}
{Birrer}, S., {Shajib}, A.~J., {Galan}, A., {et~al.} 2020, \aap, 643, A165

\bibitem[Birrer {et~al.}(2021)]{lenstro03}
Birrer, S., Shajib, A.~J., Gilman, D., {et~al.} 2021, Journal of Open Source
  Software, 6, 3283

\bibitem[{Blecher} {et~al.}(2019)]{Blecher19}
{Blecher}, T., {Deane}, R., {Heywood}, I., \& {Obreschkow}, D. 2019, \mnras,
  484, 3681

\bibitem[{Bolton} {et~al.}(2008{\natexlab{a}})]{Bolton_sample_08}
{Bolton}, A.~S., {Burles}, S., {Koopmans}, L. V.~E., {et~al.}
  2008{\natexlab{a}}, \apj, 682, 964

\bibitem[{Bolton} {et~al.}(2006{\natexlab{a}})]{lens_search_1}
{Bolton}, A.~S., {Burles}, S., {Koopmans}, L. V.~E., {Treu}, T., \&
  {Moustakas}, L.~A. 2006{\natexlab{a}}, \apj, 638, 703

\bibitem[{Bolton} {et~al.}(2006{\natexlab{b}})]{Bolton06}
{Bolton}, A.~S., {Burles}, S., {Koopmans}, L. V.~E., {Treu}, T., \&
  {Moustakas}, L.~A. 2006{\natexlab{b}}, \apj, 638, 703

\bibitem[{Bolton} {et~al.}(2006{\natexlab{c}})]{ppn_bolton06}
{Bolton}, A.~S., {Rappaport}, S., \& {Burles}, S. 2006{\natexlab{c}}, \prd, 74,
  061501

\bibitem[{Bolton} {et~al.}(2008{\natexlab{b}})]{lmass_Bolton08}
{Bolton}, A.~S., {Treu}, T., {Koopmans}, L. V.~E., {et~al.} 2008{\natexlab{b}},
  \apj, 684, 248

\bibitem[{Bolton} {et~al.}(2012)]{lmass_Bolton12}
{Bolton}, A.~S., {Brownstein}, J.~R., {Kochanek}, C.~S., {et~al.} 2012, \apj,
  757, 82

\bibitem[{Boyce} {et~al.}(2006)]{central_image_Boyce06}
{Boyce}, E.~R., {Winn}, J.~N., {Hewitt}, J.~N., \& {Myers}, S.~T. 2006, \apj,
  648, 73

\bibitem[{Brownstein} {et~al.}(2012)]{lens_search_4}
{Brownstein}, J.~R., {Bolton}, A.~S., {Schlegel}, D.~J., {et~al.} 2012, \apj,
  744, 41

\bibitem[Buchner {et~al.}(2014)]{pymultinest}
Buchner, J., Georgakakis, A., Nandra, K., {et~al.} 2014, Astronomy and
  Astrophysics, 564, A125

\bibitem[{Cabanac} {et~al.}(2007)]{lens_search_2}
{Cabanac}, R.~A., {Alard}, C., {Dantel-Fort}, M., {et~al.} 2007, \aap, 461, 813

\bibitem[{Cao} {et~al.}(2017)]{ppn_caoshuo17}
{Cao}, S., {Li}, X., {Biesiada}, M., {et~al.} 2017, \apj, 835, 92

\bibitem[{Cao} {et~al.}(2020)]{lens_search_9}
{Cao}, X., {Li}, R., {Shu}, Y., {et~al.} 2020, \mnras, 499, 3610

\bibitem[{Cappellari}(2002)]{Cappellari02}
{Cappellari}, M. 2002, \mnras, 333, 400

\bibitem[{Cappellari}(2008)]{Cappellari08}
{Cappellari}, M. 2008, \mnras, 390, 71

\bibitem[{Cappellari} {et~al.}(2013)]{Cappellari13}
{Cappellari}, M., {Scott}, N., {Alatalo}, K., {et~al.} 2013, \mnras, 432, 1709

\bibitem[{Chen} {et~al.}(2016)]{lmass_percent_1}
{Chen}, G. C.~F., {Suyu}, S.~H., {Wong}, K.~C., {et~al.} 2016, \mnras, 462,
  3457

\bibitem[{Chen} {et~al.}(2019)]{slope_chen19}
{Chen}, Y., {Li}, R., {Shu}, Y., \& {Cao}, X. 2019, \mnras, 488, 3745

\bibitem[{Cheng} {et~al.}(2020)]{Cheng20}
{Cheng}, C., {Cao}, X., {Lu}, N., {et~al.} 2020, \apj, 898, 33

\bibitem[{Collett}(2015)]{Collett15}
{Collett}, T.~E. 2015, \apj, 811, 20

\bibitem[{Collett} {et~al.}(2018)]{ppn_Collett2018}
{Collett}, T.~E., {Oldham}, L.~J., {Smith}, R.~J., {et~al.} 2018, Science, 360,
  1342

\bibitem[{Cornachione} {et~al.}(2018)]{Cornachione18}
{Cornachione}, M.~A., {Bolton}, A.~S., {Shu}, Y., {et~al.} 2018, \apj, 853, 148

\bibitem[{Crain} {et~al.}(2015)]{hydro_sim_eagle2}
{Crain}, R.~A., {Schaye}, J., {Bower}, R.~G., {et~al.} 2015, \mnras, 450, 1937

\bibitem[Diemer(2018)]{colossus}
Diemer, B. 2018, The Astrophysical Journal Supplement Series, 239, 35

\bibitem[{Ding} {et~al.}(2021)]{Ding21}
{Ding}, X., {Treu}, T., {Birrer}, S., {et~al.} 2021, arXiv:2006.08619

\bibitem[{Du} {et~al.}(2020)]{lmass_Du20}
{Du}, W., {Zhao}, G.-B., {Fan}, Z., {et~al.} 2020, \apj, 892, 62

\bibitem[{Dutton} \& {Treu}(2014)]{Dutton14}
{Dutton}, A.~A., \& {Treu}, T. 2014, \mnras, 438, 3594

\bibitem[Dye {et~al.}(2008)]{Dye08}
Dye, S., Evans, N.~W., Belokurov, V., Warren, S.~J., \& Hewett, P. 2008,
  Monthly Notices of the Royal Astronomical Society, 388, 384–392

\bibitem[{Enzi} {et~al.}(2020)]{Enzi20}
{Enzi}, W., {Vegetti}, S., {Despali}, G., {Hsueh}, J.-W., \& {Metcalf}, R.~B.
  2020, \mnras, 496, 1718

\bibitem[{Falco} {et~al.}(1985)]{Falco1985}
{Falco}, E.~E., {Gorenstein}, M.~V., \& {Shapiro}, I.~I. 1985, \apjl, 289, L1

\bibitem[Feroz {et~al.}(2009)]{multinest}
Feroz, F., Hobson, M.~P., \& Bridges, M. 2009, Monthly Notices of the Royal
  Astronomical Society, 398, 1601

\bibitem[Foreman-Mackey(2016)]{corner}
Foreman-Mackey, D. 2016, The Journal of Open Source Software, 1, 24

\bibitem[Foreman-Mackey {et~al.}(2013)]{emcee}
Foreman-Mackey, D., Hogg, D.~W., Lang, D., \& Goodman, J. 2013, Publications of
  the Astronomical Society of the Pacific, 125, 306

\bibitem[{Foreman-Mackey} {et~al.}(2013)]{Foreman-Mackey13}
{Foreman-Mackey}, D., {Hogg}, D.~W., {Lang}, D., \& {Goodman}, J. 2013, \pasp,
  125, 306

\bibitem[{Gavazzi} {et~al.}(2012)]{lens_search_3}
{Gavazzi}, R., {Treu}, T., {Marshall}, P.~J., {Brault}, F., \& {Ruff}, A. 2012,
  \apj, 761, 170

\bibitem[{Gavazzi} {et~al.}(2007)]{lmass_Gavazzi07}
{Gavazzi}, R., {Treu}, T., {Rhodes}, J.~D., {et~al.} 2007, \apj, 667, 176

\bibitem[{Genel} {et~al.}(2014)]{hydro_sim_Illustris1}
{Genel}, S., {Vogelsberger}, M., {Springel}, V., {et~al.} 2014, \mnras, 445,
  175

\bibitem[{Gilman} {et~al.}(2020)]{DM_Gilman20}
{Gilman}, D., {Birrer}, S., {Nierenberg}, A., {et~al.} 2020, \mnras, 491, 6077

\bibitem[{Gilman} {et~al.}(2019)]{DM_Gilman19}
{Gilman}, D., {Birrer}, S., {Treu}, T., {Nierenberg}, A., \& {Benson}, A. 2019,
  \mnras, 487, 5721

\bibitem[{Gomer} \& {Williams}(2021)]{Gomer21}
{Gomer}, M.~R., \& {Williams}, L. L.~R. 2021, \mnras, 504, 1340

\bibitem[{Gong} {et~al.}(2019)]{csst1}
{Gong}, Y., {Liu}, X., {Cao}, Y., {et~al.} 2019, \apj, 883, 203

\bibitem[{He} {et~al.}(2018)]{lmass_he18}
{He}, Q., {Li}, R., {Lim}, S., {et~al.} 2018, \mnras, 480, 5084

\bibitem[{He} {et~al.}(2020{\natexlab{a}})]{DM_He20}
{He}, Q., {Robertson}, A., {Nightingale}, J., {et~al.} 2020{\natexlab{a}},
  arXiv e-prints, arXiv:2010.13221

\bibitem[{He} {et~al.}(2020{\natexlab{b}})]{lmass_he20}
{He}, Q., {Li}, H., {Li}, R., {et~al.} 2020{\natexlab{b}}, \mnras, 496, 4717

\bibitem[{He} {et~al.}(2020{\natexlab{c}})]{lens_search_11}
{He}, Z., {Er}, X., {Long}, Q., {et~al.} 2020{\natexlab{c}}, \mnras, 497, 556

\bibitem[{Hernquist}(1990)]{Hernquist1990}
{Hernquist}, L. 1990, \apj, 356, 359

\bibitem[{Huang} {et~al.}(2020)]{lens_search_10}
{Huang}, X., {Storfer}, C., {Ravi}, V., {et~al.} 2020, \apj, 894, 78

\bibitem[Hunter(2007)]{matplotlib}
Hunter, J.~D. 2007, Computing in Science \& Engineering, 9, 90

\bibitem[{Ivezi{\'c}} {et~al.}(2019)]{lsst}
{Ivezi{\'c}}, {\v Z}., {Kahn}, S.~M., {Tyson}, J.~A., {et~al.} 2019, \apj, 873,
  111

\bibitem[{Keeton}(2003)]{central_image_Keeton03}
{Keeton}, C.~R. 2003, \apj, 582, 17

\bibitem[{Keeton} {et~al.}(1997)]{Keeton1997}
{Keeton}, C.~R., {Kochanek}, C.~S., \& {Seljak}, U. 1997, \apj, 482, 604

\bibitem[Kelly(2020)]{pyquad}
Kelly, A.~J. 2020, pyquad

\bibitem[{Kochanek}(2020)]{Kochanek20}
{Kochanek}, C.~S. 2020, \mnras, 493, 1725

\bibitem[{Kochanek}(2021)]{Kochanek21}
{Kochanek}, C.~S. 2021, \mnras, 501, 5021

\bibitem[{Koopmans} {et~al.}(2006{\natexlab{a}})]{lmass_Koopmans06}
{Koopmans}, L. V.~E., {Treu}, T., {Bolton}, A.~S., {Burles}, S., \&
  {Moustakas}, L.~A. 2006{\natexlab{a}}, \apj, 649, 599

\bibitem[{Koopmans} {et~al.}(2006{\natexlab{b}})]{slope_Koopmans06}
{Koopmans}, L. V.~E., {Treu}, T., {Bolton}, A.~S., {Burles}, S., \&
  {Moustakas}, L.~A. 2006{\natexlab{b}}, \apj, 649, 599

\bibitem[{Koopmans} {et~al.}(2009)]{slope_Koopmans09}
{Koopmans}, L.~V.~E., {Bolton}, A., {Treu}, T., {et~al.} 2009, \apjl, 703, L51

\bibitem[Lam {et~al.}(2015)]{numba}
Lam, S.~K., Pitrou, A., \& Seibert, S. 2015, Proceedings of the Second Workshop
  on the LLVM Compiler Infrastructure in HPC - LLVM '15, 1

\bibitem[{Li} {et~al.}(2018{\natexlab{a}})]{Hongyu18}
{Li}, H., {Mao}, S., {Cappellari}, M., {et~al.} 2018{\natexlab{a}}, \mnras,
  476, 1765

\bibitem[{Li} {et~al.}(2016)]{DM_RanLi16}
{Li}, R., {Frenk}, C.~S., {Cole}, S., {et~al.} 2016, \mnras, 460, 363

\bibitem[{Li} {et~al.}(2017)]{DM_RanLi17}
{Li}, R., {Frenk}, C.~S., {Cole}, S., {Wang}, Q., \& {Gao}, L. 2017, \mnras,
  468, 1426

\bibitem[{Li} {et~al.}(2018{\natexlab{b}})]{slope_lirui18}
{Li}, R., {Shu}, Y., \& {Wang}, J. 2018{\natexlab{b}}, \mnras, 480, 431

\bibitem[{Li} {et~al.}(2019)]{Ran19}
{Li}, R., {Li}, H., {Shao}, S., {et~al.} 2019, \mnras, 490, 2124

\bibitem[{Li} {et~al.}(2020)]{lens_search_12}
{Li}, R., {Napolitano}, N.~R., {Tortora}, C., {et~al.} 2020, \apj, 899, 30

\bibitem[Lin(2018)]{pynufft}
Lin, J.~M. 2018, Journal of Imaging, 4, 1

\bibitem[{Mao} \& {Schneider}(1998)]{DM_Mao1998}
{Mao}, S., \& {Schneider}, P. 1998, \mnras, 295, 587

\bibitem[{Maresca} {et~al.}(2021)]{unphy_src_Maresca21}
{Maresca}, J., {Dye}, S., \& {Li}, N. 2021, arXiv:2012.04665

\bibitem[{Marinacci} {et~al.}(2018)]{hydro_sim_TNG1}
{Marinacci}, F., {Vogelsberger}, M., {Pakmor}, R., {et~al.} 2018, \mnras, 480,
  5113

\bibitem[{Marques-Chaves} {et~al.}(2020)]{Marques20}
{Marques-Chaves}, R., {{\'A}lvarez-M{\'a}rquez}, J., {Colina}, L., {et~al.}
  2020, \mnras, 499, L105

\bibitem[Meneghetti {et~al.}(2013)]{Meneghetti_2013}
Meneghetti, M., Bartelmann, M., Dahle, H., \& Limousin, M. 2013, Space Science
  Reviews, 177, 31–74

\bibitem[{Millon} {et~al.}(2020)]{Millon20}
{Millon}, M., {Galan}, A., {Courbin}, F., {et~al.} 2020, \aap, 639, A101

\bibitem[Miranda(2018)]{pyswarms}
Miranda, L. J.~V. 2018, Journal of Open Source Software, 3

\bibitem[{Mukherjee} {et~al.}(2018)]{Mukherjee18}
{Mukherjee}, S., {Koopmans}, L. V.~E., {Metcalf}, R.~B., {et~al.} 2018, \mnras,
  479, 4108

\bibitem[{Naiman} {et~al.}(2018)]{hydro_sim_TNG2}
{Naiman}, J.~P., {Pillepich}, A., {Springel}, V., {et~al.} 2018, \mnras, 477,
  1206

\bibitem[{Navarro} {et~al.}(1996)]{nfw_profile}
{Navarro}, J.~F., {Frenk}, C.~S., \& {White}, S. D.~M. 1996, \apj, 462, 563

\bibitem[{Nelson} {et~al.}(2015)]{hydro_sim_Illustris3}
{Nelson}, D., {Pillepich}, A., {Genel}, S., {et~al.} 2015, Astronomy and
  Computing, 13, 12

\bibitem[{Nelson} {et~al.}(2018)]{hydro_sim_TNG3}
{Nelson}, D., {Pillepich}, A., {Springel}, V., {et~al.} 2018, \mnras, 475, 624

\bibitem[Newman {et~al.}(2013)]{AndrewNewman2013}
Newman, A.~B., Treu, T., Ellis, R.~S., {et~al.} 2013, The Astrophysical
  Journal, 765, 24

\bibitem[{Newton} {et~al.}(2011)]{Newton11}
{Newton}, E.~R., {Marshall}, P.~J., {Treu}, T., {et~al.} 2011, \apj, 734, 104

\bibitem[{Nightingale} \& {Dye}(2015)]{Nightingale15}
{Nightingale}, J.~W., \& {Dye}, S. 2015, \mnras, 452, 2940

\bibitem[{Nightingale} {et~al.}(2018)]{Nightingale18}
{Nightingale}, J.~W., {Dye}, S., \& {Massey}, R.~J. 2018, \mnras, 478, 4738

\bibitem[Nightingale {et~al.}(2021{\natexlab{a}})]{pyautofit}
Nightingale, J.~W., Hayes, R.~G., \& Griffiths, M. 2021{\natexlab{a}}, Journal
  of Open Source Software, 6, 2550

\bibitem[{Nightingale} {et~al.}(2019)]{lmass_Nightingale19}
{Nightingale}, J.~W., {Massey}, R.~J., {Harvey}, D.~R., {et~al.} 2019, \mnras,
  489, 2049

\bibitem[Nightingale {et~al.}(2021{\natexlab{b}})]{pyautolens}
Nightingale, J.~W., Hayes, R.~G., Kelly, A., {et~al.} 2021{\natexlab{b}},
  Journal of Open Source Software, 6, 2825

\bibitem[O’Riordan {et~al.}(2020)]{Riordan2020}
O’Riordan, C.~M., Warren, S.~J., \& Mortlock, D.~J. 2020, Monthly Notices of
  the Royal Astronomical Society

\bibitem[{Paraficz} {et~al.}(2016)]{lens_search_7}
{Paraficz}, D., {Courbin}, F., {Tramacere}, A., {et~al.} 2016, \aap, 592, A75

\bibitem[Pedregosa {et~al.}(2011)]{scikit-learn}
Pedregosa, F., Varoquaux, G., Gramfort, A., {et~al.} 2011, Journal of Machine
  Learning Research, 12, 2825

\bibitem[{Petrillo} {et~al.}(2017)]{lens_search_8}
{Petrillo}, C.~E., {Tortora}, C., {Chatterjee}, S., {et~al.} 2017, \mnras, 472,
  1129

\bibitem[{Pillepich} {et~al.}(2018)]{hydro_sim_TNG4}
{Pillepich}, A., {Nelson}, D., {Hernquist}, L., {et~al.} 2018, \mnras, 475, 648

\bibitem[{Planck Collaboration} {et~al.}(2016)]{planck15}
{Planck Collaboration}, {Ade}, P.~A.~R., {Aghanim}, N., {et~al.} 2016, \aap,
  594, A13

\bibitem[{Price-Whelan} {et~al.}(2018)]{astropy2}
{Price-Whelan}, A.~M., {Sip{\H{o}}cz}, B.~M., {G{\"u}nther}, H.~M., {et~al.}
  2018, \aj, 156, 123

\bibitem[{Quinn} {et~al.}(2016)]{central_image_Quinn16}
{Quinn}, J., {Jackson}, N., {Tagore}, A., {et~al.} 2016, \mnras, 459, 2394

\bibitem[{Ravasi} \& {Vasconcelos}(2020)]{pylops}
{Ravasi}, M., \& {Vasconcelos}, I. 2020, SoftwareX, 11, 100361

\bibitem[{Refsdal}(1964)]{Refsdal1964}
{Refsdal}, S. 1964, \mnras, 128, 307

\bibitem[{Ritondale} {et~al.}(2019{\natexlab{a}})]{Ritondale19}
{Ritondale}, E., {Auger}, M.~W., {Vegetti}, S., \& {McKean}, J.~P.
  2019{\natexlab{a}}, \mnras, 482, 4744

\bibitem[{Ritondale} {et~al.}(2019{\natexlab{b}})]{DM_Ritondale19}
{Ritondale}, E., {Vegetti}, S., {Despali}, G., {et~al.} 2019{\natexlab{b}},
  \mnras, 485, 2179

\bibitem[{Rizzo} {et~al.}(2020)]{Rizzo20}
{Rizzo}, F., {Vegetti}, S., {Powell}, D., {et~al.} 2020, \nat, 584, 201

\bibitem[{Ruff} {et~al.}(2011)]{slope_ruff11}
{Ruff}, A.~J., {Gavazzi}, R., {Marshall}, P.~J., {et~al.} 2011, \apj, 727, 96

\bibitem[{Rusin} \& {Ma}(2001)]{central_image_Rusin01}
{Rusin}, D., \& {Ma}, C.-P. 2001, \apjl, 549, L33

\bibitem[{Schaye} {et~al.}(2015)]{hydro_sim_eagle1}
{Schaye}, J., {Crain}, R.~A., {Bower}, R.~G., {et~al.} 2015, \mnras, 446, 521

\bibitem[{Schneider} {et~al.}(1992)]{lensing_textbook}
{Schneider}, P., {Ehlers}, J., \& {Falco}, E.~E. 1992, {Gravitational Lenses}

\bibitem[{Schneider} \& {Sluse}(2013)]{Schneider_Sluse_13}
{Schneider}, P., \& {Sluse}, D. 2013, \aap, 559, A37

\bibitem[{Schneider} \& {Sluse}(2014)]{Schneider_Sluse_14_SPT}
{Schneider}, P., \& {Sluse}, D. 2014, \aap, 564, A103

\bibitem[{Schwab} {et~al.}(2010)]{ppn_Schwab_2010}
{Schwab}, J., {Bolton}, A.~S., \& {Rappaport}, S.~A. 2010, \apj, 708, 750

\bibitem[{S{\'e}rsic}(1963)]{sersic1963}
{S{\'e}rsic}, J.~L. 1963, Boletin de la Asociacion Argentina de Astronomia La
  Plata Argentina, 6, 41

\bibitem[{Shajib}(2019)]{new_guass_decomposition}
{Shajib}, A.~J. 2019, \mnras, 488, 1387

\bibitem[{Shajib} {et~al.}(2020)]{pure_lensing_Shajib20}
{Shajib}, A.~J., {Treu}, T., {Birrer}, S., \& {Sonnenfeld}, A. 2020, arXiv
  e-prints, arXiv:2008.11724

\bibitem[{Shu} {et~al.}(2016{\natexlab{a}})]{lens_search_5}
{Shu}, Y., {Bolton}, A.~S., {Kochanek}, C.~S., {et~al.} 2016{\natexlab{a}},
  \apj, 824, 86

\bibitem[{Shu} {et~al.}(2016{\natexlab{b}})]{Shu16}
{Shu}, Y., {Bolton}, A.~S., {Mao}, S., {et~al.} 2016{\natexlab{b}}, \apj, 833,
  264

\bibitem[{Shu} {et~al.}(2017)]{lens_search_6}
{Shu}, Y., {Brownstein}, J.~R., {Bolton}, A.~S., {et~al.} 2017, \apj, 851, 48

\bibitem[{Sonnenfeld}(2018)]{stellar_dynamics_H0_analysis}
{Sonnenfeld}, A. 2018, \mnras, 474, 4648

\bibitem[{Sonnenfeld} {et~al.}(2013)]{slope_Sonnenfeld13}
{Sonnenfeld}, A., {Treu}, T., {Gavazzi}, R., {et~al.} 2013, \apj, 777, 98

\bibitem[Speagle(2020)]{dynesty}
Speagle, J.~S. 2020, Monthly Notices of the Royal Astronomical Society, 493,
  3132

\bibitem[{Springel} {et~al.}(2018)]{hydro_sim_TNG5}
{Springel}, V., {Pakmor}, R., {Pillepich}, A., {et~al.} 2018, \mnras, 475, 676

\bibitem[{Suyu} {et~al.}(2010)]{Suyu10}
{Suyu}, S.~H., {Marshall}, P.~J., {Auger}, M.~W., {et~al.} 2010, \apj, 711, 201

\bibitem[{Suyu} {et~al.}(2009)]{Suyu09}
{Suyu}, S.~H., {Marshall}, P.~J., {Blandford}, R.~D., {et~al.} 2009, \apj, 691,
  277

\bibitem[{Suyu} {et~al.}(2006)]{Suyu06}
{Suyu}, S.~H., {Marshall}, P.~J., {Hobson}, M.~P., \& {Blandford}, R.~D. 2006,
  \mnras, 371, 983

\bibitem[{Suyu} {et~al.}(2013)]{Suyu13}
{Suyu}, S.~H., {Auger}, M.~W., {Hilbert}, S., {et~al.} 2013, \apj, 766, 70

\bibitem[{Suyu} {et~al.}(2017)]{lmass_percent_2}
{Suyu}, S.~H., {Bonvin}, V., {Courbin}, F., {et~al.} 2017, \mnras, 468, 2590

\bibitem[{Tagore} {et~al.}(2018)]{Tagore18}
{Tagore}, A.~S., {Barnes}, D.~J., {Jackson}, N., {et~al.} 2018, \mnras, 474,
  3403

\bibitem[{Tagore} \& {Keeton}(2014)]{Tagore14}
{Tagore}, A.~S., \& {Keeton}, C.~R. 2014, \mnras, 445, 694

\bibitem[{Tessore} \& {Metcalf}(2015)]{Tessore15}
{Tessore}, N., \& {Metcalf}, R.~B. 2015, \aap, 580, A79

\bibitem[{Treu} {et~al.}(2006)]{lmass_Treu06}
{Treu}, T., {Koopmans}, L.~V., {Bolton}, A.~S., {Burles}, S., \& {Moustakas},
  L.~A. 2006, \apj, 640, 662

\bibitem[{Van de Vyvere} {et~al.}(2020)]{boundary_shear_effect}
{Van de Vyvere}, L., {Sluse}, D., {Mukherjee}, S., {Xu}, D., \& {Birrer}, S.
  2020, \aap, 644, A108

\bibitem[{van der Walt} {et~al.}(2011)]{numpy}
{van der Walt}, S., {Colbert}, S.~C., \& {Varoquaux}, G. 2011, Computing in
  Science Engineering, 13, 22

\bibitem[Van~der Walt {et~al.}(2014)]{scikit-image}
Van~der Walt, S., Sch{\"o}nberger, J.~L., Nunez-Iglesias, J., {et~al.} 2014,
  PeerJ, 2, e453

\bibitem[Van~Rossum \& Drake(2009)]{python}
Van~Rossum, G., \& Drake, F.~L. 2009, Python 3 Reference Manual (Scotts Valley,
  CA: CreateSpace)

\bibitem[{Vegetti} \& {Koopmans}(2009)]{lmass_Vegetti09}
{Vegetti}, S., \& {Koopmans}, L.~V.~E. 2009, \mnras, 392, 945

\bibitem[{Vegetti} {et~al.}(2014)]{DM_Vegetti14}
{Vegetti}, S., {Koopmans}, L.~V.~E., {Auger}, M.~W., {Treu}, T., \& {Bolton},
  A.~S. 2014, \mnras, 442, 2017

\bibitem[{Vegetti} {et~al.}(2010)]{Vegetti10}
{Vegetti}, S., {Koopmans}, L.~V.~E., {Bolton}, A., {Treu}, T., \& {Gavazzi}, R.
  2010, \mnras, 408, 1969

\bibitem[{Vegetti} {et~al.}(2012{\natexlab{a}})]{DM_Vegetti12}
{Vegetti}, S., {Lagattuta}, D.~J., {McKean}, J.~P., {et~al.}
  2012{\natexlab{a}}, \nat, 481, 341

\bibitem[{Vegetti} {et~al.}(2012{\natexlab{b}})]{Vegetti12}
{Vegetti}, S., {Lagattuta}, D.~J., {McKean}, J.~P., {et~al.}
  2012{\natexlab{b}}, \nat, 481, 341

\bibitem[{Virtanen} {et~al.}(2020)]{scipy}
{Virtanen}, P., {Gommers}, R., {Oliphant}, T.~E., {et~al.} 2020, Nature
  Methods, 17, 261

\bibitem[{Vogelsberger} {et~al.}(2014)]{hydro_sim_Illustris2}
{Vogelsberger}, M., {Genel}, S., {Springel}, V., {et~al.} 2014, \nat, 509, 177

\bibitem[{Wang} {et~al.}(2019)]{Wang19}
{Wang}, Y., {Vogelsberger}, M., {Xu}, D., {et~al.} 2019, \mnras, 490, 5722

\bibitem[{Warren} \& {Dye}(2003)]{Warren03}
{Warren}, S.~J., \& {Dye}, S. 2003, \apj, 590, 673

\bibitem[{Winn} {et~al.}(2004)]{central_image_Winn04}
{Winn}, J.~N., {Rusin}, D., \& {Kochanek}, C.~S. 2004, \nat, 427, 613

\bibitem[{Wong} {et~al.}(2020)]{wong20}
{Wong}, K.~C., {Suyu}, S.~H., {Chen}, G. C.~F., {et~al.} 2020, \mnras, 498,
  1420

\bibitem[{Wyithe} {et~al.}(2001)]{gnfw_Wyithe}
{Wyithe}, J.~S.~B., {Turner}, E.~L., \& {Spergel}, D.~N. 2001, \apj, 555, 504

\bibitem[{Xu} {et~al.}(2016)]{Xu16}
{Xu}, D., {Sluse}, D., {Schneider}, P., {et~al.} 2016, \mnras, 456, 739

\bibitem[{Yang} {et~al.}(2021)]{Yang21}
{Yang}, L., {Roberts-Borsani}, G., {Treu}, T., {et~al.} 2021, \mnras, 501, 1028

\bibitem[{Yang} {et~al.}(2020)]{ppn_yang2020}
{Yang}, T., {Birrer}, S., \& {Hu}, B. 2020, \mnras, 497, L56

\bibitem[{Y{\i}ld{\i}r{\i}m} {et~al.}(2020)]{next_gen_td2020}
{Y{\i}ld{\i}r{\i}m}, A., {Suyu}, S.~H., \& {Halkola}, A. 2020, \mnras, 493,
  4783

\bibitem[{Zhang} {et~al.}(2007)]{central_image_Zhang07}
{Zhang}, M., {Jackson}, N., {Porcas}, R.~W., \& {Browne}, I.~W.~A. 2007,
  \mnras, 377, 1623

\bibitem[{Zhao}(1996)]{gnfw_Zhao1996}
{Zhao}, H. 1996, \mnras, 278, 488

\end{thebibliography}

\appendix
\section{Ancillary lens properties}
\label{sec:appdix_lens_prop}
\begin{figure}
	\includegraphics[width=\columnwidth]{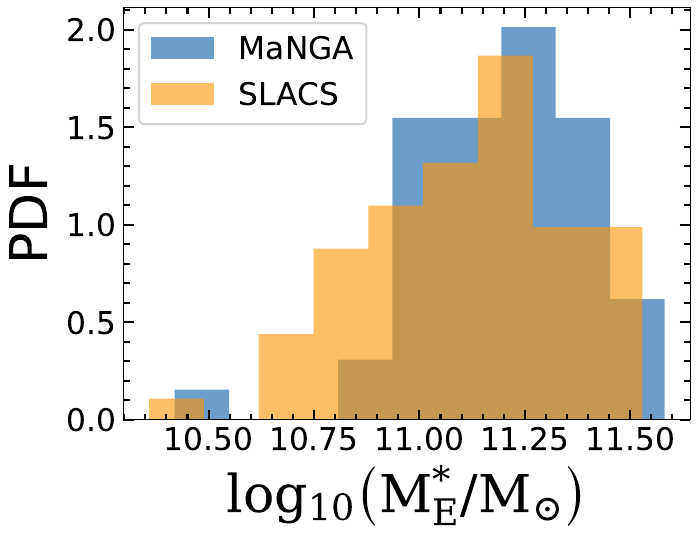}
    \caption{\reply{the PDF of the projected stellar mass within the Einstein radius of the lens galaxy. For both real SLACS lenses (orange) and our mock MaNGA lenses (blue).}
    }
    \label{fig:apend_lens_prop_1}
\end{figure}
\reply{
Figure~\ref{fig:apend_lens_prop_1} shows the projected stellar mass within the Einstein radius for both real SLACS lenses and our mock MaNGA lenses. The stellar mass of the SLACS data is from \citet{Auger09_stellar}, derived by the stellar population synthesis model assuming the Salpeter IMF. We find the stellar mass distribution of our mock MaNGA lenses is also similar to SLACS lenses.
}

\begin{figure}
	\includegraphics[width=\columnwidth]{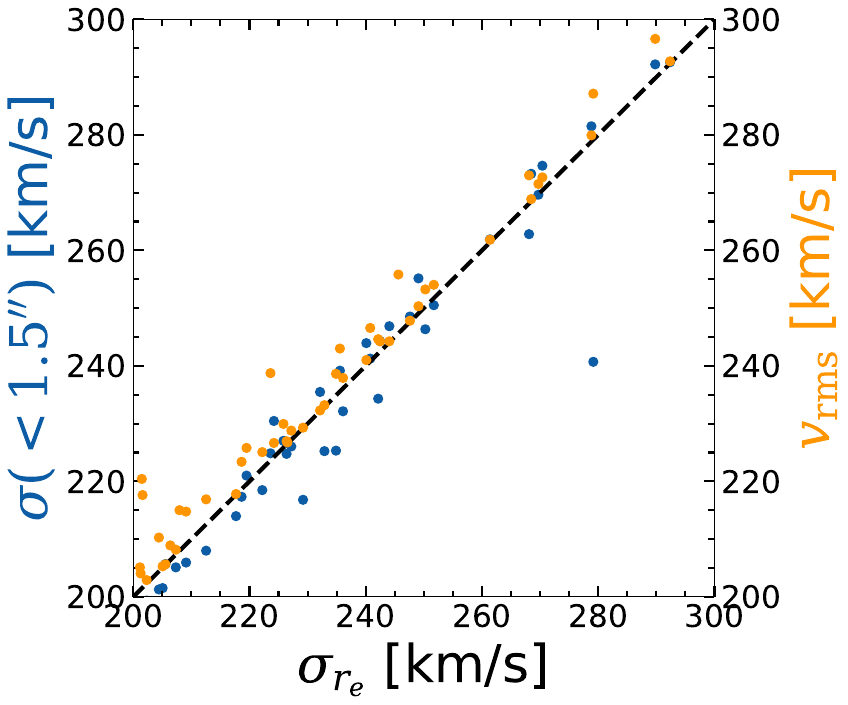}
    \caption{\reply{Blue dots show a comparison between the average velocity dispersion within the half-light radius ($\sigma_{r_e}^2$) and that within the 1.5\arcsec ($\sigma(<1.5^{\arcsec})$) for our mock MaNGA lenses. The relationship between the average second velocity moment within the half-light radius ($v_{\rm rms}$) and $\sigma_{r_e}$ is shown with the orange dots. $v_{\rm rms}$ is defined as $v_{\rm rms} = \sqrt{v_{\rm los}^2+\sigma_{r_e}^2}$, where $v_{\rm los}$ represents the line-of-sight velocity. We can find $v_{\rm rms}\approx \sigma_{r_e}$, which means our mock lenses are made up of dispersion dominated ETGs.}
}
    \label{fig:apend_lens_prop_2}
\end{figure}
\reply{
In Figure~\ref{fig:apend_lens_prop_2}, we use the blue dots to show the relationship between the luminosity-weighted velocity dispersion within the half-light radius ($\sigma_{r_e}$) and that within the 1.5\arcsec (the aperture size of the SDSS fiber, denoted as $\sigma(<1.5^{\arcsec})$. We find those two values agree broadly, which indicates we could use $\sigma_{r_e}$ as a proxy to select SLACS-like ETGs. We don’t include any galaxy rotation requirement for ETGs when generating mock lenses. The orange dots in Figure~\ref{fig:apend_lens_prop_2} show the relationship between the luminosity-weighted second velocity moment within $r_e$ ($v_{\rm rms}$) and $\sigma_{r_e}$. $v_{\rm rms}$ is defined as $v_{\rm rms} = \sqrt{v_{\rm los}^2+\sigma_{r_e}^2}$, where $v_{\rm los}$ represents the line-of-sight velocity. We can find $v_{\rm rms}\approx \sigma_{r_e}$, which means our mock lenses are made up of dispersion dominated ETGs.
}

\section{\reply{Deflection angle and lensing potential calculation}}
\label{sec:appdix_defl}
\begin{figure}
	\includegraphics[width=\columnwidth]{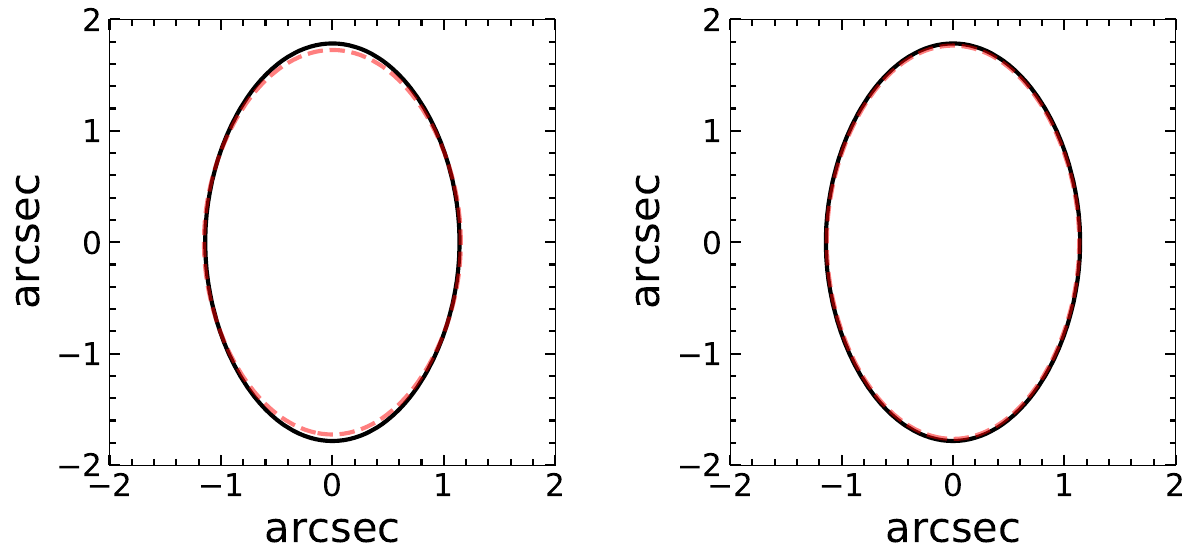}
    \caption{The critical line of an ideal EPL lens. Left panel: the black line is the critical line draw based on the deflection angle map that is given by the analytical formula of the EPL model, which represents the true critical line;  while the red dashed line is the critical line draw based on the deflection angle map which is calculated numerically. Right panel: similar to the left panel, the only difference is the convergence map used to numerically calculate the deflection angle map has a larger size (\reply{$40^{\arcsec}\times40^{\arcsec}$} compare to $5^{\arcsec}\times5^{\arcsec}$ in the left panel).}
    \label{fig:apend_a_1}
\end{figure}

In this work, we calculate the deflection angle map numerically from the convergence map. \citet{boundary_shear_effect} shows that the truncation of the convergence map changes the symmetry of the lens's density distribution at the boundary, which may result in ``pseudo shear''. To avoid this boundary effect, we need to expand the size of the convergence map when we calculate the deflection angle numerically. In Figure~\ref{fig:apend_a_1}, \reply{we use an ideal EPL lens (slope=2.1, axis-ratio=0.6) to demonstrate this effect}. The black line in the left panel is the critical line drawn based on the deflection angle given by the analytical formula of the EPL model, while the red dashed line is the critical line drawn based on the deflection angle field that is calculated numerically (with convergence map size $5^{\arcsec}\times5^{\arcsec}$). It is clear the red dashed line does not coincide with the black line, which proves that the truncation of the convergence map brings significant ``pseudo shear''; The right panel is the calculation result when the convergence map size is \reply{$40^{\arcsec}\times40^{\arcsec}$} (the scheme shown in Section~\ref{sec:gen_data}). We find that the red line coincides with the black line, therefore the boundary effect can be ignored for this case.

\begin{figure}
	\includegraphics[width=\columnwidth]{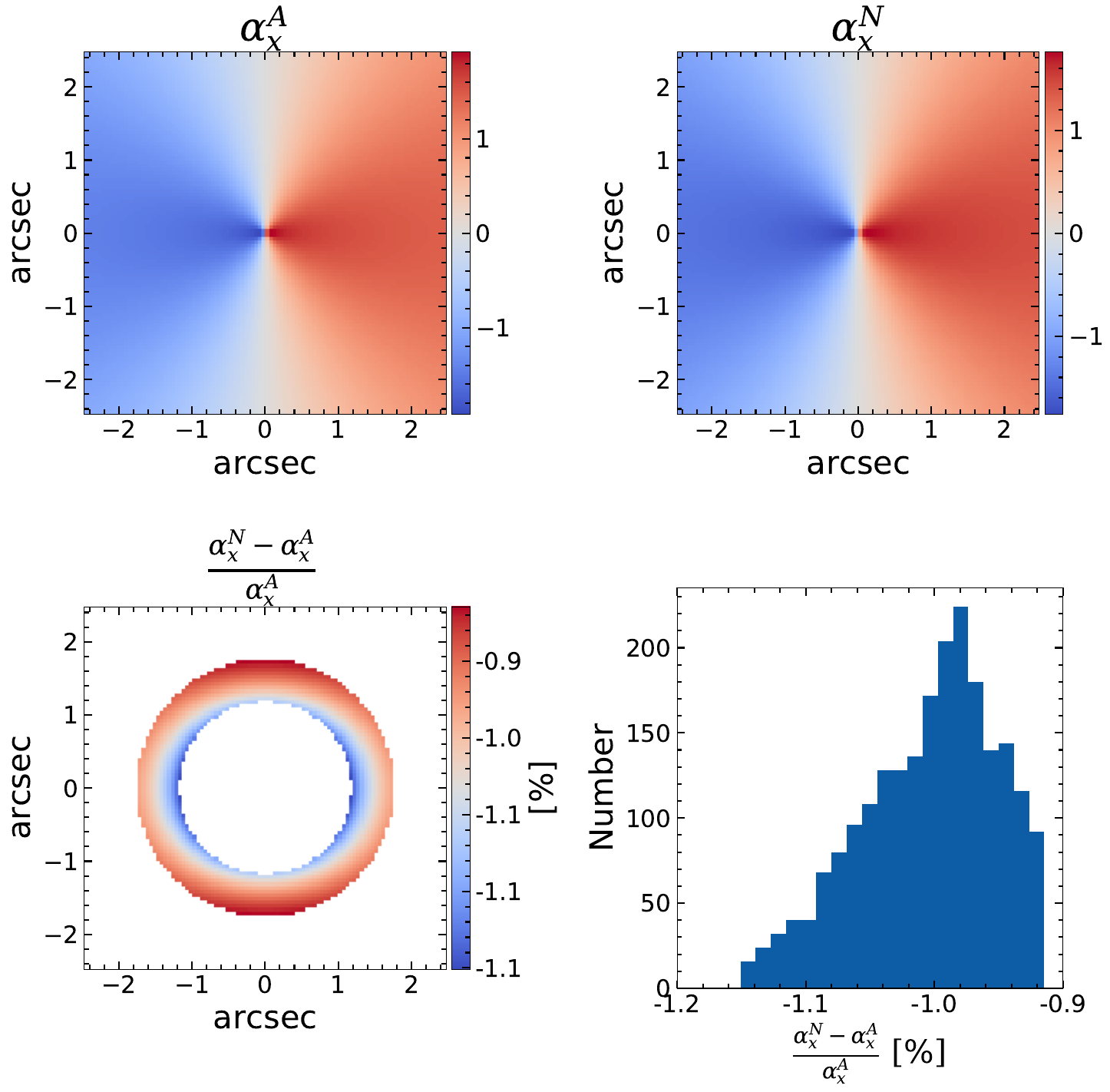}
    \caption{\reply{Top-left: the x-component of the deflection angle calculated according to the analytical formula of the EPL model ($\alpha_x^{A}$). Top-right: the x-component of the deflection angle calculated from the convergence map using the numerical scheme shown in Section~\ref{sec:gen_data} ($\alpha_x^{N}$). Bottom-left: the relative error ($\frac{\alpha_x^N-\alpha_x^A}{\alpha_x^A}$) of the numerically calculated deflection angle in the region where the lensing signal appears ([$0.8\times\theta_{\rm E}$, $1.2\times\theta_{\rm E}$]). Bottom-right: histogram of the relative error shown in the bottom-left panel.}}
    \label{fig:apend_a_2}
\end{figure}

\begin{figure}
	\includegraphics[width=\columnwidth]{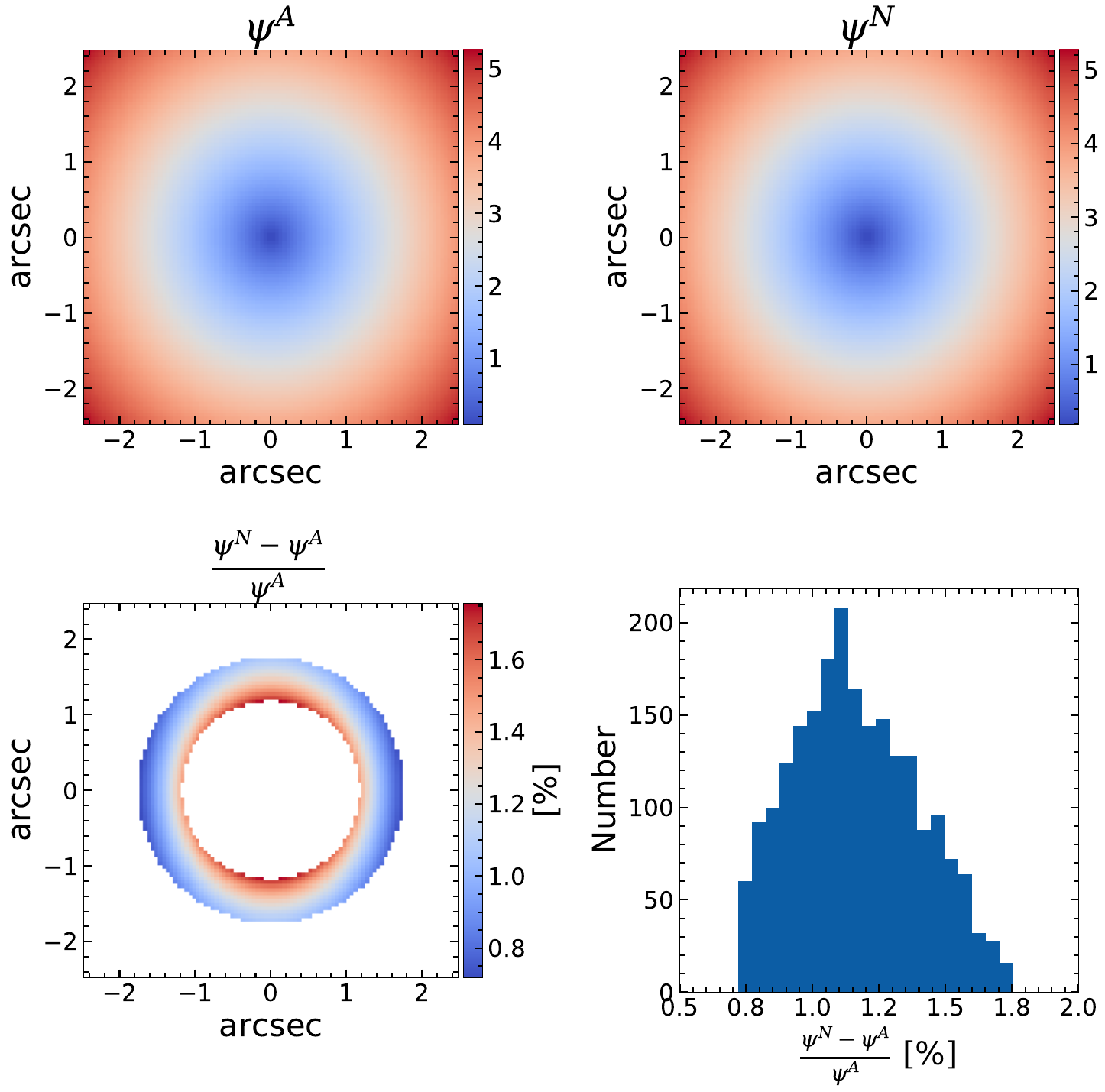}
    \caption{\reply{Similar to figure~\ref{fig:apend_a_2}, the relative error of the numerically calculated lensing potential ($\frac{\psi^N-\psi^A}{\psi^A}$).}}
    \label{fig:apend_a_3}
\end{figure}

\reply{In Figures~\ref{fig:apend_a_2} and \ref{fig:apend_a_3}, we use an ideal EPL mass model to test the numerical error of the deflection angle and lensing potential calculated from the convergence. The parameter values of EPL model are set to the typical values of our simulated lens samples (slope = 2.1, Einstein radius = 1.5\arcsec, axis-ratio = 0.8). We find that the numerical scheme described in Section~\ref{sec:gen_data} produces a numerical error of $\sim 1\%$ for the deflection angle and lensing potential in the annular region where the lensing signal appears. Since our mock lensing images are generated from the numerically calculated deflection angles; the numerical errors we present here mean the lens galaxies in our mock data actually approach slightly biased real MaNGA ETGs. 
Using a higher level of oversampling (criterion-1 in Section~\ref{sec:gen_data}), and increasing the size of the convergence map (criterion-2 in Section~\ref{sec:gen_data}) can further reduce the numerical error of the deflection angle and lensing potential calculation. However, this requires more computational resources (currently, it takes $\sim 20$ hours to calculate the convergence, deflection angle and lensing potential for each lens). The faster algorithm proposed in \citet{new_guass_decomposition} may be used in future to overcome the numerical calculation errors due to limited computational resources.}

\section{Estimate the error of \texorpdfstring{$\gamma_{\rm ppn}$}{Lg}}
% https://tex.stackexchange.com/questions/53513/hyperref-token-not-allowed/53514
\label{sec:ppn_err}
\reply{
Equation~\eqref{eq:ppn} is equivalent to 
\begin{equation}
\left(\frac{\theta_E^{\rm dyn}}{\theta_E^{\rm lensing}}\right)^2 = \frac{1+\gamma_{\rm ppn}}{2}~,
\label{eq:ppn_err_1}
\end{equation}
where $\theta_E^{\rm dyn}$ is the Einstein radius derived from dynamical mass measurements, which is assumed to be the input true value ($\theta_E^{\rm true}$), presumably the mass measurements of stellar dynamics are error-free. From our lens modeling results, we had estimated the relative measurement error (statistical + systematic) on Einstein radius is $\sim$0.1\%, therefore we have,
\begin{equation}
\frac{\theta_E^{\rm lensing}-\theta_E^{\rm true}}{\theta_E^{\rm true}} \equiv 
\frac{\theta_E^{\rm lensing}-\theta_E^{\rm dyn}}{\theta_E^{\rm dyn}}  = 
\frac{\theta_E^{\rm lensing}}{\theta_E^{\rm dyn}} - 1 \sim \mathcal{N}(0,\,\sigma^{2})|_{\sigma=0.1\%}\,
\label{eq:ppn_err_2}
\end{equation}
where $\mathcal{N}(0,\,\sigma^{2})$ represents the normal distribution with mean $0$ and variance $\sigma^2$. Equation~\eqref{eq:ppn_err_2} indicates $\frac{\theta_E^{\rm lensing}}{\theta_E^{\rm dyn}}$ follows the distribution $\mathcal{N}(1,\,(0.1\%)^{2})$. 

To derive the error of $\gamma_{\rm ppn}$, we resort to the Monte Carlo way. We generate a set of discrete $\frac{\theta_E^{\rm lensing}}{\theta_E^{\rm dyn}}$ samples follow the distribution we derived above. For each sample, we calculate the corresponding  $\gamma_{\rm ppn}$ value with equation~\eqref{eq:ppn_err_1}. Eventually, The error of $\gamma_{\rm ppn}$ is given by the standard deviation value of $\gamma_{\rm ppn}$ samples.
}
\section{Non-MST lenses}
\label{sec:appdix_nmst}
For most outliers (9 of 11) whose physical parameters are significantly misestimated, we find their existence can be understood via the Mass-Sheet Transformation.  In this section, we present the remaining two outliers which may reflect the more complex lensing degeneracy.

\begin{figure}
\includegraphics[width=\columnwidth]{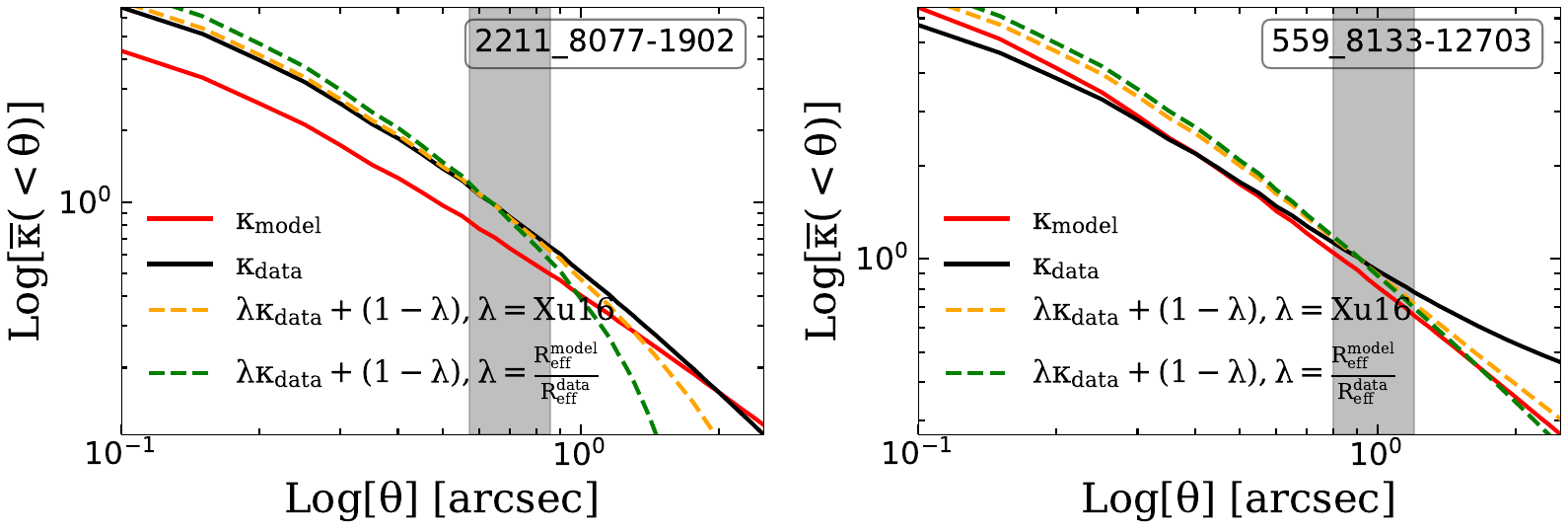}
\caption{Similar to Figure~\ref{fig:7_outliers_avg_kappa_prof}, averaged radial profile of $\rm \bar{\kappa}(<\theta)$ for lens system ``2211\_8077-1902'' (left-panel) and ``559\_8133-12703'' (right-panel).}
\label{fig:nmst_1}
\end{figure}

\begin{figure}
\includegraphics[width=\columnwidth]{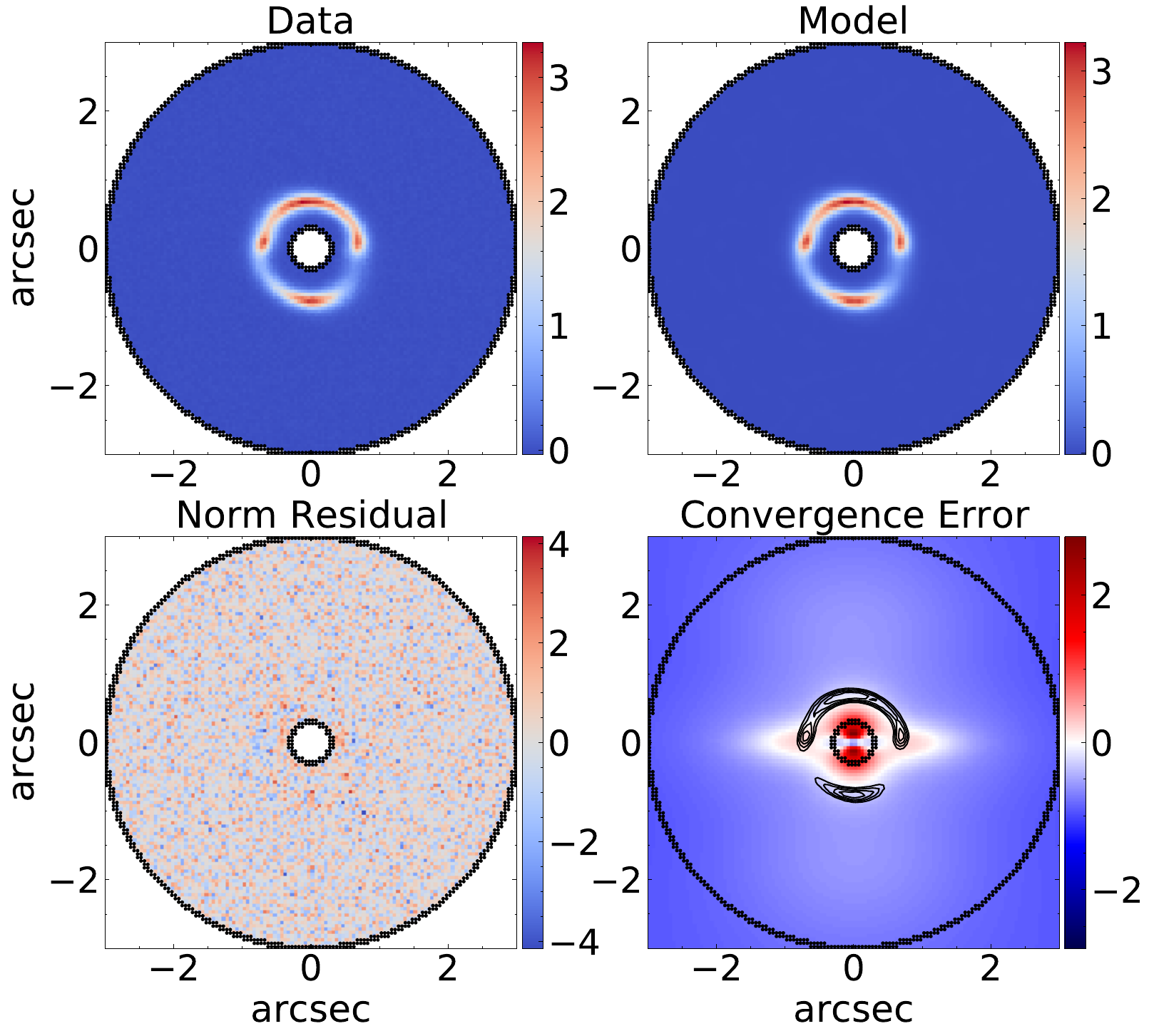}
\caption{Similar to Figure~\ref{fig:4_typical_case_model_result}, the best-fit image of the lens system ``2211\_8077-1902''. We can see from the bottom-right panel that the relative error map is not elliptical-symmetric, which indicates a significant angular mismatch between the model and data exists. }
\label{fig:nmst_2}
\end{figure}

\begin{figure}
\includegraphics[width=\columnwidth]{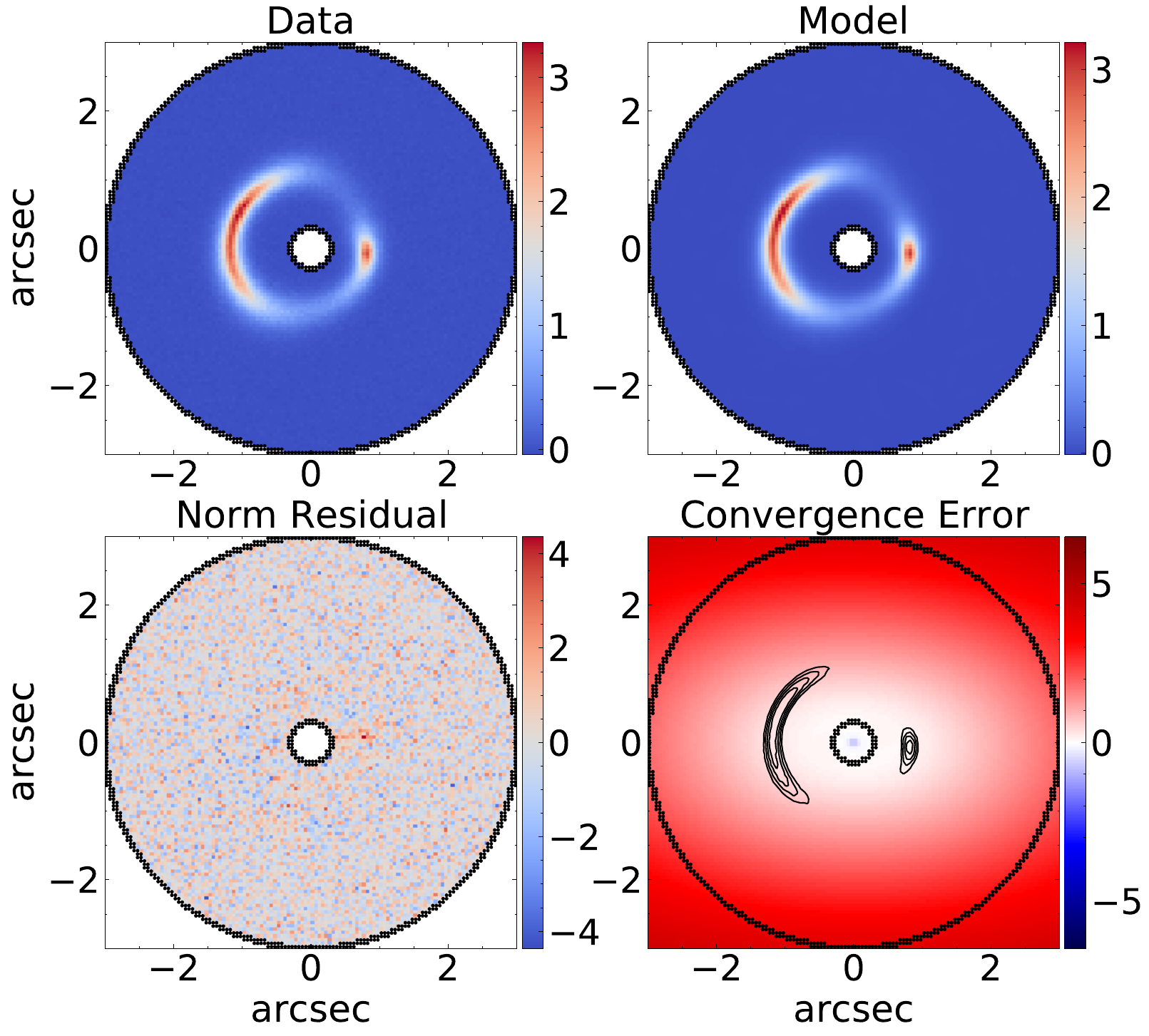}
\caption{Similar to Figure~\ref{fig:nmst_2}, the best-fit image of the lens system ``559\_8133-12703''.}
\label{fig:nmst_3}
\end{figure}

Figure~\ref{fig:nmst_1} shows the azimuthal averaged radial profile of $\rm \bar{\kappa}(<\theta)$ for lens system ``2211\_8077-1902'' and ``559\_8133-12703''. It is clearly shown that the red line, green line, and orange line do not coincide so that the MST explanation discussed in Section~\ref{sec:mass_slope} is not applicable. We further compare the deflector's mass distribution predicted by the lens model (EPL + shear mass assumption) with that of ground truth value, we find the difference between those two mass distributions (bottom-right panel in Figure~\ref{fig:nmst_2} and \ref{fig:nmst_3}) have significant angular structure particularly in the region where the extended arc appears. Therefore, a uniform mass sheet is not sufficient to compensate for the mass distribution mismatch between the ideal EPL+shear model and our mock lenses. The degeneracy in these two lens systems manifests the general Souce-Position Transformation.

\section{$H_0$ from the relative time delay}
\label{sec:h0_td}

The relative time delay defined in equation~\eqref{eq:rel_td} can be abbreviated as
\begin{equation}
\Delta t_{AB} = \frac{\rm \eta}{H_0}~,
\label{eq:h0_td_1}
\end{equation}
where $\rm \eta$ is a factor only depending on the lens and source redshift, the position of images A and B, and corresponding lensing potential values. 

For  our mock ``MaNGA lenses'',  suppose the true relative time delay between an image pair is given as
\begin{equation}
\Delta t^D = \frac{\rm  \eta^D}{H^D_0}~,
\label{eq:h0_td_2}
\end{equation}
Since our image-based lens modeling can not perfectly reconstruct the lensing potential at the location of the image pair, $\rm \eta^M$ is a biased estimation of $\rm \eta^D$, hence our model prediction to the relative time delay $\Delta t^M$ is also biased, i.e.
\begin{equation}
\Delta t^M = \frac{\rm  \eta^M}{H^D_0}~.
\label{eq:h0_td_3}
\end{equation}
Assume the time-domain observation can perfectly measure the true relative time delay ($\Delta t^D$),  Then we can use $\Delta t^D$ and the lensing potential measured from the lens model ($\rm \eta^M$) to infer $H_0$,
\begin{equation}
H^M_0 = \frac{\rm  \eta^M}{\Delta t^D},
\label{eq:h0_td_4}
\end{equation}
insert equations~\eqref{eq:h0_td_3} into \eqref{eq:h0_td_4} to arrive at 
\begin{equation}
\begin{aligned}
H^M_0 &= \frac{\rm  \eta^M}{\Delta t^D} \\
&= \left(\frac{\Delta t^\mathrm{M}}{\Delta t^\mathrm{D}}\right) H_0^\mathrm{D}~.
\end{aligned}
\label{eq:h0_td_5}
\end{equation}
Therefore, our biased measurement of the relative time delay can be directly interpreted as a biased measurement of $H_0$ via equation~\eqref{eq:h0_td_5}.

\label{lastpage}
\end{document}